\documentclass{IEEEtran}
 \pdfoutput=1 
 \usepackage{amsmath,amssymb,amsfonts}
\usepackage{algorithm}
\usepackage{algorithmicx}
\usepackage{algpseudocode}
\usepackage{graphicx}
\usepackage{textcomp}
\usepackage{subcaption}
\usepackage{multicol}
\usepackage{lipsum}
\usepackage{enumerate}
\usepackage{bm}
\usepackage[numbers]{natbib}
\newtheorem{lemma}{Lemma}

\newtheorem{corollary}{Corollary}
\newtheorem{definition}{Definition}

\newcommand{\RNum}[1]{\uppercase\expandafter{\romannumeral #1\relax}}

\def\BibTeX{{\rm B\kern-.05em{\sc i\kern-.025em b}\kern-.08em
    T\kern-.1667em\lower.7ex\hbox{E}\kern-.125emX}}
\begin{document}

\title{Self-Stabilizing Periodic Mutual-exclusive Propagation in Sparse Networks
\thanks{This work has been submitted to the IEEE for possible publication. Copyright may be transferred without notice, after which this version may no longer be accessible.}
}

\author{Shaolin Yu*, Jihong Zhu, Jiali Yang\\Tsinghua University, Beijing, China \\ysl8088@163.com}

\maketitle

\begin{abstract}
Message propagation is fundamental in constructing distributed systems upon sparsely connected communication networks.
For providing easy message propagation primitives, the mutual-exclusive propagation (MEP) of one-bit messages is investigated with a simplified discrete system model.
Inspired by natural propagation systems, efficient self-stabilizing periodic MEP processes are proposed with the MEP systems.
For handling the worst-case scenarios, the MEP systems are generally analyzed with formal proofs upon arbitrarily connected networks.
It shows that the MEP systems can be stabilized with arbitrary initial system states within a short bounded time.
Meanwhile, the numeric simulation shows that the propagation errors can be significantly reduced if the message delays are randomly distributed.
Propagation patterns are also discussed in further deriving finer propagation results in the MEP systems.
It shows that bounded clock drifts and some benign faults can be well handled in the MEP systems.
Featured by its very simple mechanism, the proposed MEP primitive can be employed as a practical building block of upper-layer self-stabilizing synchronous systems.
\end{abstract}


\section{Introduction}
\label{sec:Introduction}
Propagation is widely investigated in both nature and artificial systems.
For example, the scintillations of stars observed on the ground are explained with the propagation of lights in irregular media \citep{Uscinski1977}.
The perceptions of humans and machines usually rely on propagations of some excitation signals in neural networks and their simplified simulations.
Physically, for deducing any information from observations of remotely happened events, it is critical to make reasonable assumptions for the related signal propagations.
Especially in distributed systems, the properties of message propagations (or saying information diffusions~\citep{Cristian1995Atomic,diffusion1998Gargano}), which include message correctness, message consistency, message delays, message lengths, message cycles, etc., are often the basis for designing real-world distributed protocols.
In general, the more constraints can be imposed on the related propagation primitives, the easier the distributed protocols would be derived with the information deduced from local observations of the distributed components.

Unfortunately, the costs of the propagation primitives often rise sharply in realizing the desired constraints in real-world systems.
For example, although global physical broadcasts and peer-to-peer authenticated messages are highly desired in synchronization~\citep{Fine2002Elson,FTSP2004}, Byzantine agreement~\citep{Dolev1983Authenticated,SrikanthSimulating}, self-stabilizing protocols~\citep{Daliot2003inspired,Daliot2005stabilization}, and so on, the costs of realizing reliable broadcast and message authentication are often prohibitively high.
Especially, as the network connectivity required in Byzantine fault-tolerant message primitives \citep{SrikanthSimulating, Dolev1982StrikeAgain} is gravely bounded, these theoretically high-reliable message primitives can hardly be well supported in most real-world low-cost communication systems.
In providing relatively more accessible but still useful propagation primitives upon arbitrarily connected networks, \cite{YuCOTS2021} proposes a practical message propagation primitive in building self-stabilizing synchronization systems without introducing high extra costs.
In \cite{YuCOTS2021}, the propagation of the messages is governed by the so-called \emph{mutual-exclusive} propagation rule.
Namely, when this propagation rule is imposed on some specific messages, these messages would be propagated in the network without being overlapped with each other.
This is a suitable property when the system scale is large and only simple local algorithms can be afforded in the distributed computing and communicating components.
Also, with the self-stabilization property, higher-layer self-stabilizing protocols can be directly built upon this basic mutual-exclusive propagation (MEP) primitive.

However, the MEP process introduced in \cite{YuCOTS2021} is tightly coupled with other stages of the self-stabilizing synchronization protocol.
Especially in providing the desired self-stabilization property, some time parameters of the MEP primitive are related to the overall synchronization cycles.
Meanwhile, the overall synchronization process proposed in \cite{YuCOTS2021} relies on high-precision delay measurement and global message broadcast, which can hardly be supported in large scale systems with limited power supplies such as the wireless sensor networks (WSN), Internet of things (IoT), and other emerging low-cost networking technologies.
Also, the MEP patterns introduced in \cite{YuCOTS2021} are not well investigated in sparsely connected networks.

In this paper, we develop the MEP systems in an intuitive and self-contained way.
Firstly, inspired by the natural wave propagations governed by the reaction-diffusion rules in excitable media, we would develop the MEP systems with a simplified discrete dynamic model.
With this model, we extend the basic one-shot MEP (o-MEP) to periodic MEP (p-MEP).
Then, we would introduce the self-stabilizing p-MEP (sp-MEP) system, which can be employed as a general building block in constructing self-stabilizing distributed protocols in real-time systems with arbitrarily connected communication networks.
We would see that, just like controlling the strokes in combustion engines, the self-stabilization of the self-contained MEP systems can be naturally supported by properly configuring some specific time parameters.
Especially, we would give a finer analysis of the time parameters of the self-contained sp-MEP systems.
Then, we would also discuss the propagation patterns of sp-MEP upon some specific sparsely connected networks.
For the ease of system realization, only $1$-bit messages are employed in the discussed MEP systems without any real-time delay measurement or message broadcast.

The remainder of this paper is organized as follows.
In section~\ref{sec:Related}, some existing propagation primitives are briefly reviewed.
In section~\ref{sec:Models}, the basic system model and the related propagation problems are intuitively proposed.
In section~\ref{sec:MEP}, the basic properties of the sp-MEP systems are formally investigated.
In section~\ref{sec:Pattern}, we study the propagation patterns and investigate some MEP systems with numeric simulations in some sparsely connected networks.
Lastly, the paper is concluded in section~\ref{sec:Conclusion}.

\section{Related works}
\label{sec:Related}
In practice, almost all \emph{distributed} systems are developed relying on some kinds of signal propagations.
For example, receiver-to-receiver broadcast synchronization relies on some physically broadcasted beacon signals in providing the desired high precision~\citep{Fine2002Elson}.
End-to-end messages employed in Internet applications are propagated under some packet-routing rules~\citep{CK1974Protocol,mills1991internet}.
Point-to-point secure communication protocols proposed in sparsely connected networks are built upon low-layer redundant transmission schemes~\citep{Dwork1986,Upfal1992,Chandran2010}.

Generally, we can coarsely categorize the various propagation primitives by the types of the propagated signals.
At the lowest level, physical propagation primitives often rely on physical properties of the propagation media~\citep{Fine2002Elson}.
For a typical example, traditional bidirectional half-duplex Ethernet and wireless Ethernet are built upon physical signals that can be freely propagated in the shared media.
With the properties of the shared media, the propagated signals can often pass through each other without being mutually exclusive.
As a result, these signals might collide in the receiving devices, which is often handled by high-layer protocols.
For another example, some classical cellular automata \citep{Schiff2007Cellular} are built upon excitation signals (can be viewed as $1$-bit messages) propagated in excitable media upon specific discrete networks.
In this case, the properties of the propagation media largely depend on the excitation rules realized in the mass-deployed cellular units.
Analogously, by viewing the discrete networks and the excitation rules as the propagation spaces (can be with arbitrary topologies) and the propagation laws (can be any computable functions), respectively, almost any imaginable and useful continuous propagation processes can be discretely simulated in cellular automata.
However, in considering the implementation of the cellular units, the propagation laws should be as easy as possible, drawing the real-world limitation over the power of large-scale cellular automata, especially in the presence of faults.
In 1956, Von Neumann~\citep{Neumann1956} first attacked the fault-tolerance problem with single output automata (or saying the \emph{basic organs}).
By interconnecting the basic organs with wires (or \emph{nerve bundles}) in some specific ways, \cite{Neumann1956} first shows that a number of random faults (being under a small threshold) generated in the basic organs (and the input signals) in \emph{intuitionistic logics} can be well-tolerated in the randomly multiplexed systems with only $1$-bit messages and simple logic operations.
However, in tolerating such random faults, the activation of the overall system should be driven by a global clock that generates periodic common pulse signals for all basic organs.

Later in \cite{Cristian1995Atomic,diffusion1998Gargano}, several asynchronous information diffusion primitives are proposed for reaching atomic broadcasts in peer-to-peer networks.
By propagating messages with bounded delays, \cite{Cristian1995Atomic} shows that atomic broadcasts can be well-simulated in arbitrarily connected networks in considering the worst-case delays.
In \cite{diffusion1998Gargano} the communication complexity of fault-tolerant information diffusion is further investigated, and an optimal cost fault-tolerant broadcasting algorithm is proposed.
However, these information diffusion primitives all employ multi-valued messages and asynchronous message diffusion processes, with which the worst-case delays (in considering message flooding, message queueing and routing, media accessing, etc.) can be high in networks of large diameters.
In \cite{Bickson2008Gaussian}, Gaussian belief propagation is investigated in handling large-scale distributed computing problems with synchronous and asynchronous algorithms.
In practice, however, only synchronous belief propagation algorithms are implemented, requiring the communication networks to be synchronized first.
Although asynchronous propagation algorithms are also investigated in \cite{Bickson2008Gaussian}, the implementation of efficient and reliable large-scale asynchronous communication systems is still not an easy task.
Other asynchronous information diffusion algorithms go even more complex and diversified as being developed in modern large-scale social networks \citep{info8040118}.
However, the stabilization properties of the various asynchronous information diffusion systems are not fully understood, especially in considering the \emph{immune} nodes can become susceptible again.

In considering the needs of real-world high-reliable applications, people also try to build more robust systems with complicated distributed components in relatively small-scale systems.
In this approach, various message propagation primitives are proposed as building blocks of distributed computing systems.
For example, instead of directly building Byzantine agreement systems from scratch, authenticated message broadcast primitives can be simulated first in peer-to-peer message-passing networks \citep{SrikanthSimulating}.
Upon this, some high-layer Byzantine fault-tolerant protocols can be efficiently built-in some easy-understood ways \citep{Toueg1987Fast,Daliot2006Agreement}.
In the same vein, high-layer self-stabilizing protocols can be built upon a self-stabilizing Byzantine agreement which in turn be built upon some low-layer self-stabilizing propagation primitives \citep{Daliot2005stabilization,Lenzen2019Almost}.
However, most of these propagation primitives require high network connectivity of the distributed systems.
In sparsely connected networks, several secure communication protocols are built upon specific message transmission schemes, which can be employed as propagation primitives in higher-layer fault-tolerant protocols~\citep{Dwork1986,Upfal1992,Chandran2010}.
However, as the higher-layer primitives would become more complex in tolerating malign faults of the distributed components, the cost of realizing the \emph{correct} distributed components required in these systems can often skyrocket.

In practice, for building efficient multi-valued message-passing systems, it is often desired that some underlying synchronous systems be established first, with which the higher-layer messages can be synchronously delivered in peer-to-peer networks.
For this, the MEP primitive is proposed in \cite{YuCOTS2021} with employing $1$-bit synchronization signals for reaching efficient self-stabilizing synchronization and TDMA communication in arbitrarily connected networks.
In the provided prototype system, the MEP primitive is realized with COTS Ethernet components and other low-cost common devices.
However, the propagation primitive proposed in \cite{YuCOTS2021} is tightly coupled with the overall synchronization algorithm that relies upon the global broadcast of the multi-valued messages.

For large-scale networks, an important issue is to avoid multi-valued messages and expensive broadcast.
Meanwhile, as the initial states of the distributed components in large-scale networks can hardly be pre-configured, the self-contained MEP systems are also expected to be self-stabilizing, i.e., always converge into some desired system states in a bounded time from arbitrary initial system states.
Furthermore, when the MEP system is stabilized, it is desired that the accumulated message delays can be further lowered, and thus more time resources (such as the time-slots in TDMA communication) can be reserved for high-layer applications.
As far as we know, these problems are not well handled in existing propagation primitives.

\section{System models and the problems}
\label{sec:Models}

\subsection{Mutual-exclusive propagation system}
The MEP system $\mathcal{S}$ consists of $n$ basic propagation units (or saying cells, denoted as $N$) which are arbitrarily connected in the bidirectional propagation network $G=(N,E)$.
To be reactive in the MEP system, each cell $i\in N$ has an excitation state $x_i\in \{0,1\}$ at any given instant $t\in \mathbb T$, where $\mathbb T\subseteq\mathbb R$ is the real time set.
Denoting $x(i,t)$ as the value of $x_i$ at $t$, $x(i,t)=1$ holds iff (if and only if) $i$ is in its excited state at $t$.
In the MEP system $\mathcal{S}$, $x_i$ can toggle its value (from $0$ to $1$ or from $1$ to $0$) at $t$ with the following MEP rules.

\begin{enumerate}[R1)]
\item \label{enum_r1}
if $\forall t_1\in [t-\tau_1,t):x(i,t_1)=0$ holds for some $\tau_1$, $i$ would be externally triggered and $x_i$ would toggle from $0$ to $1$ at $t$.
\item \label{enum_r2}
if $\forall t_1\in [t-\tau_0,t):x(i,t_1)=1$ holds for some $\tau_0$, $i$ would be restored and $x_i$ would toggle from $1$ to $0$ at $t$.
\item \label{enum_r3}
if $\forall t_1\in [t-\tau',t):x(i,t_1)=0$ holds and $i$ remotely observes the trigger events of a sufficient number of its adjacent cells during $[t-\tau',t]$ with some $\tau'$, $i$ would be internally triggered (in $G$) and $x_i$ would toggle from $0$ to $1$ at $t$.
\end{enumerate}

For remotely observing the instantaneous trigger events, when a cell $i\in N$ is triggered at $t$, $i$ would emit trigger signals to be propagated to the neighbors of $i$ (denoted as $N_i$) upon $G$.
For analysis, the signal delay of the trigger signal emitted from the cell $i$ to the cell $j\in N_i$ is defined as the sum of all delays experienced by this specific trigger signal (including the emission delay, physical propagation delay, receiving delay, signal processing delay, etc.).
In considering imperfect real-world propagation processes, when $i$ is triggered, the trigger signals propagated to different neighbors of $i$ can experience different signal delays.
Nevertheless, following the bounded-delay model \citep{DolevPulseBoundedDelay2007}, the signal delays of the trigger signals between two adjacent cells are assumed to be bounded, providing that no fault occurs.
Meanwhile, for approximately measuring the real time, each nonfaulty cell $i\in N$ is equipped with a local clock $c_i$ with bounded drift rates.
Concretely, we say the cells in $Q\subseteq N$ are nonfaulty iff the signal delays and drift rates of the local clocks generated in these cells are bounded in $[0,d]$ and $[-\rho,\rho]$, respectively.
For simplicity, we assume that all cells in the propagation network are nonfaulty in the basic system model.
Then, in discussing the fault-tolerance problem, the extended system model would also cover some restricted failure modes of the faulty cells.

\subsection{Inspiration from the natural world}

For grasping the basic idea of MEP, an intuitive analogy can be the propagation in some excitable media \citep{winfree1980geometry, Bernus2015excitable}.
For a concrete example, the propagation of flare waves in some combustible substances can usually be described with the following reaction-diffusion equation~\citep{LAKSHMIKANTHAM1981243}
\begin{eqnarray}
\label{eq:rde}
\left\{
\begin{aligned}
&\frac{\partial u}{\partial t} = K_1\Delta u+Qve^{-\frac{E}{Ru}} \\
&\frac{\partial v}{\partial t} = K_2\Delta v-ve^{-\frac{E}{Ru}}
\end{aligned}
\right.
\end{eqnarray}
where $\Delta$ is the Laplace operator, $u$ is the temperature, $v$ is the combustible concentration, and the other quantities are some constant numbers.
Naturally, under the constraint of (\ref{eq:rde}), a sufficiently strong initial combustion can increase the temperature of the surrounding combustible substances with the temperature being diffused from the hotter area to the colder ones.
With this and sufficiently high combustible concentration, some chemical reactions can occur to increase the temperature in these adjacent areas to induce further chemical reactions.
Meanwhile, the chemical reaction will cause a decrease in the combustible concentration, along with the flammable substances being diffused from higher concentration areas to the lower ones.

Despite this relatively convoluted relation, a simple result of the reaction-diffusion processes can be observed from real-world combustion engines.
In combustion engines, the periodic combustion strokes can be well controlled.
This is also the desired property of the MEP systems.

\subsection{A discrete system representation}
Following the heuristic work of Alan Turing \citep{TURING1990153}, with a finite number of \emph{cells} and the propagation network $G$, here we can also construct a discrete MEP system analogous to (\ref{eq:rde}).
Concretely, since the propagation of the messages can be more naturally described with discrete instantaneous events, the propagation constraint can be simplified as
\begin{eqnarray}
\label{eq:sig_propa}
\left\{
\begin{aligned}
&u(i,t) = (w(i,t) + \sum_{j\in N_i}u^{(i)}(j,t))\cdot  v(i,t^-) \\
&v(i,t) = \prod_{t_0\in [t-\tau,t]}\lnot u(i,t_0)
\end{aligned}
\right.
\end{eqnarray}
where
 $u(i,t)$ is the trigger signal (whether a trigger event occurs) of the cell $i$ at $t$,
 $v(i,t)$ is the propagability (whether new trigger event is allowed) of $i$ at $t$,
 $w(i,t)$ is the external trigger signal (whether an external trigger event occurs) of $i$ at $t$,
 $u^{(i)}(j,t)$ is the trigger signal of $j$ observed in cell $i$ at $t$,
 and $+$, $\cdot$, $\sum$, and $\prod$ are boolean operations commonly defined upon $\{0,1\}$.
For simplicity, here we only consider the basic propagation rule with which a cell would be internally triggered iff $v(i,t^-)=1$ and $\sum_{j\in N_i}u^{(i)}(j,t)>0$, where $t^-$ is earlier than $t$ with an infinitesimal.

In considering real-world realization with non-ignorable clock drifts, $\tau$ is bounded within $[\tau_0/(1+\rho),\tau_0/(1-\rho)]$ in (\ref{eq:sig_propa}), where $\tau_0$ is the local restoration threshold.
Meanwhile, in bounded-delay networks, $u^{(i)}(j,t)$ can be represented as
\begin{equation}\label{eq:observe_j_i}
u^{(i)}(j,t)=\sum_{ d_0=d^{(i)}(j,t-d_0)} u(j,t-d_0 )
\end{equation}
where $d^{(i)}(j,t)\in [0,d]$ is the signal delay of the trigger signal emitted in cell $j$ at $t$ to cell $i$.
With $\tau_1>\tau_0/(1-\rho)$, $u(i,t)$ can be further simplified as
\begin{eqnarray}
\label{eq:sig_propa_u1}
u(i,t) = w(i,t) + v(i,t^-)\cdot \sum_{j\in N_i}u^{(i)}(j,t)
\end{eqnarray}

Here, to investigate the self-stabilization problem in real-time systems, we still use the real time set $\mathbb T$.
For simplicity but without loss of generality, we assume that only a finite number of discrete events can occur in any finite period.
With this, we assume that the trigger event only occurs in cell $i$ when $u(i,t)=1$, denoted as the event $\mathtt{e_u}(i,t)$.
Similarly, the external trigger event (the $w$ event) only occurs in $i$ when $w(i,t)=1$, denoted as the event $\mathtt{e_w}(i,t)$.
In this way, $u(i,t)$ and $w(i,t)$ can only take the value $1$ at a finite number of discrete instants during any finite period.

This simplified MEP system $\mathcal{S}$ can be analogous to some reaction-diffusion system where the propagation area is confined with a cold boundary, the reactions occur instantaneously (infinitely fast), and the diffusions are with bounded delays.
Meanwhile, the $w$ events can be viewed as some instantaneous external reactions at somewhere of the boundary, just like the sparks generated around the valves of the combustion engines.
It should be noted that the condition $v(i,t)=1$ can be satisfied and maintained in $i$ if no reaction occurs in $i$ for a sufficiently long time.
This can be analogous to the diffusion of the new intakes of combustible substances.
To be compatible with the MEP rules, we have $v(i,t)=1-x(i,t)$.
\subsection{Basic definitions}
Generally,  a propagation occurred in $\mathcal{S}$ system, denoted as $\mathcal{P}$, can be represented by a sequence of trigger events.
According to (\ref{eq:sig_propa}), if a trigger event $\mathtt{e_u}(i,t)$ occurs, there can only be two cases.
Firstly, if $w(i,t)=0$, then $i$ is triggered with remotely observing at $t$ that some neighbor $j\in N_i$ being triggered, i.e., $u^{(i)}(j,t)=1$ holds for some $j$.
Denoting $S(i,t)=\{j\in N_i\mid u^{(i)}(j,t)=1\}$, $j\in S(i,t)$ is called the pioneer of $i$ (denoted as $h_i=j$) for $\mathtt{e_u}(i,t)$ if $|S(i,t)|=1$.
For simplicity, if $|S(i,t)|>1$, we can always choose one cell $j\in S(i,t)$ (for example choosing the cell with the smallest identifer) to be the pioneer of $i$ for $\mathtt{e_u}(i,t)$ and say $\mathtt{e_u}(i,t)$ is produced by $u^{(i)}(j,t)$.
In this case, the trigger event $\mathtt{e_u}(i,t)$ can be represented as an edge $(j,i)\in E$ of $G$ (if ignoring the time dimension).
Otherwise, if $w(i,t)=1$, then we have $S(i,t)=\{i\}$ and $i$ is the pioneer of $i$ for $\mathtt{e_u}(i,t)$ (denoted as $h_i=i$).
In this case, $\mathtt{e_u}(i,t)$ can be represented as the vertex $i\in N$ of $G$ and we say $\mathtt{e_u}(i,t)$ is produced by $w(i,t)$ (and $\mathtt{e_w}(i,t)$).
Thus, with ignoring the time dimension, $\mathcal{P}$ can also be represented by a set of directed paths on $G$.
For convenience, we say the trigger signal emitted by the pioneer of $i$ would be accepted by $i$.
Meanwhile, we say a trigger signal observed in $i$ would be rejected by $i$ with $u$ if $i$ is not restored from being excited with its latest trigger event $u$.

With the analogy of the combustion processes, it is common sense that if all cells in $N$ are restored, and there is no trigger signal being propagated in $G$ before the occurrence of $\mathcal{P}$, then each directed path in $\mathcal{P}$ would be with a finite length.
Further, if the changes from $v(i,t)=0$ to $v(i,t)=1$ are forbidden (imagining that no intake of new combustible substances), then after the occurrence of $\mathcal{P}$, each cell $i\in N$ corresponds to a unique path $\gamma_i=(i_0,i_1,\dots,i_m=i)\in \mathcal{P}$ in which $i_{k}$ is the pioneer of $i_{k+1}$ with $k\in\{0,1,\dots,m-1\}$ and $i_0$ is the pioneer of $i_0$ for the corresponding trigger events in $\mathcal{P}$.
In this case, we can reduce the directed paths in $\mathcal{P}$ to the paths $\gamma_i$ for all $i\in N$ (referred to as the propagation paths, denoted as $\gamma(\mathcal{P})=\{\gamma_i\mid i\in N\}$) and can safely denote the first vertex (referred to as the source cell) of each such path $\gamma$ as $s(\gamma)$.
With this, the desired one-shot MEP can be defined in the way following \cite{YuCOTS2021}.
\begin{definition}\citep{YuCOTS2021}
\label{def_mutex_propagation}
$\mathcal{P}$ is a one-shot MEP (o-MEP) on $G$, denoted as $\mathcal{P}\in \mathbb {M}_G$, iff for $\gamma(\mathcal{P})=\{\gamma_i\mid i\in N\}$ the following conditions hold
\begin{eqnarray}
\label{eq:def_valid} \textit{Valid:}&& \forall i\in N: s(\gamma_{i})\in N_{\mathtt{s}} \\
\label{eq:def_simple} \textit{Simple:}&& \forall \gamma\in \mathcal{P}: \forall i_1,i_2\in \gamma: i_1\neq i_2 \\
\label{eq:def_complete} \textit{Complete:}&& \forall i\in N: \exists \gamma\in \mathcal{P} :i\in \gamma \\
\label{eq:def_exclusive} \textit{Exclusive:}&& \forall i_1,i_2\in N: s(\gamma_{i_1})\neq s(\gamma_{i_2}) \to \nonumber\\
&&\gamma_{i_1} \cap \gamma_{i_2}=\emptyset \\
\label{eq:def_propagative} \textit{Propagative:}&& \forall \gamma_1,\gamma_2\in \mathcal{P}: (\gamma_1 \cap \gamma_2 \neq\emptyset \to \exists \gamma,\gamma_1',\gamma_2': \nonumber\\
&&\gamma_1=\gamma\gamma_1' \land \gamma_2=\gamma\gamma_2' \land \gamma_1'\cap \gamma_2'=\emptyset)
\end{eqnarray}
\end{definition}
Here, the set of cells that are allowed to be externally triggered is denoted as $N_{\mathtt{s}}$.
In the core discussion here, we assume $N_{\mathtt{s}}=N$.
For completeness, when $\gamma_{i}$ has no source cell (this is possible when $i$ is never triggered or $\gamma_{i}$ is infinitely long, see below), we denote $s(\gamma_{i})=\infty$.
Meanwhile, it should be noted that the \emph{Complete} property given in (\ref{eq:def_complete}) seems redundant since $\gamma(\mathcal{P})$ is represented as the set of $\gamma_i$ for all $i\in N$.
It becomes necessary when we further represent the propagation with the spanning trees.
Namely, as the o-MEP forms a directed forest in which every cell in $N$ appears and only appears once, we can partition the cells to different spanning trees according to such $\mathcal{P}$.
For the desired propagation, the cells in each such spanning tree form a propagation region, and $N$ can be partitioned into one or several propagation regions.

In the above definition of the o-MEP, we have ignored the time dimension and focused on the result of some desired propagation at some desired instant.
Generally, we can use $\gamma(\mathcal{P}(t))=\{\gamma_i(t)\mid i\in N\}$ to represent the current propagation $\mathcal{P}(t)$ at any given instant $t$, where $\gamma_i(t)=(\dots,i)$ is the \emph{latest} propagation path connected to $i$ at instant $t$.
In this general case, $\gamma_i(t)$ can be with an infinite length and can be an empty path if $i$ is never triggered.
Denoting $T_i$ as the set of all trigger instants (the instants at which some trigger events occur) of cell $i$ since $t_0$, $\gamma_i(t)$ can only be changed at the instants in $T_i$ when $t\geqslant t_0$.
Denoting $T_N=\bigcup_{i\in N}T_i$, if
\begin{eqnarray}
\label{eq:propa_separation}
\forall t_1,t_2\in T_N: |t_1-t_2|\leqslant \tau_\pi \lor |t_1-t_2|>\tau_\Delta
\end{eqnarray}
holds with some $\tau_\Delta>3\tau_\pi$, we say that $T_N$ is well-separated with $(\tau_\pi,\tau_\Delta)$ (also referred to as $\pi/\Delta$-precedence in ~\cite{Kopetz2011Principles} ).
Obviously, when $T_N$ is well-separated with $(\tau_\pi,\tau_\Delta)$, we can use a set of disjoint time intervals
\begin{eqnarray}
\label{eq:I_N}
\mathcal{I}_N=\{[t_1,t_2]\mid t_1,t_2\in T_N\land t_2-t_1\in [0,\tau_\pi]\land \nonumber\\
\forall t\in T_N: t\in [t_1,t_2]\lor t\notin [t_1-\tau_\Delta,t_2+\tau_\Delta]\}
\end{eqnarray}
to cover all instants in $T_N$.
With this, denoting the propagation process $\mathcal{P}([t_0,t])=\{ \mathtt{e_u}(i,t') \mid i\in N \land t'\in [t_0,t] \}$, if $\mathcal{P}([t_0,+\infty))$ further satisfies
\begin{eqnarray}
\label{eq:propa_separation_mutex}
\forall [t_1,t_2]\in \mathcal{I}_N:\mathcal{P}([t_1,t_2])\in \mathbb { M}_G
\end{eqnarray}
the propagation process $\mathcal{P}([t_0,+\infty))$ is called a $(\tau_\pi,\tau_\Delta)$ separated MEP process.
Furthermore, if the $(\tau_\pi,\tau_\Delta)$ separated propagation process $\mathcal{P}([t_0,+\infty))$ satisfies the following liveness constraint
\begin{eqnarray}
\label{eq:propa_separation_mutex_liveness}
\forall t\in T_N: (t,t+\tau_\nabla]\cap T_N\ne \emptyset
\end{eqnarray}
we say that $\mathcal{P}([t_0,+\infty))$ is $(\tau_\pi,\tau_\Delta,\tau_\nabla)$-periodic and the sp-MEP system $\mathcal{S}$ is stabilized since $t_0$.

\subsection{Problems}
With the intuition of the combustion engines, it is not hard to see that the desired o-MEP can be formed when the initial condition and the time parameters of the system can be well configured.
However, providing that the initial state of the system is arbitrarily configured by some malicious adversary, some cells might be in the excited state, and some others might not.
In this situation, the trigger signals emitted from one or more cells may not be able to be propagated to all cells in $G$.

\begin{figure}[htbp]
\centering
\begin{subfigure}{.23\textwidth}
\centering\includegraphics[width=1.4in]{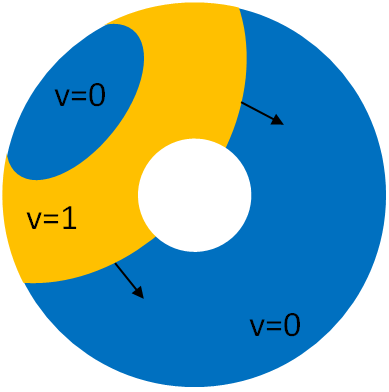}
\caption{The simplest MEP}
\label{fig:propagation0}
\end{subfigure}
\begin{subfigure}{.23\textwidth}
\centering\includegraphics[width=1.4in]{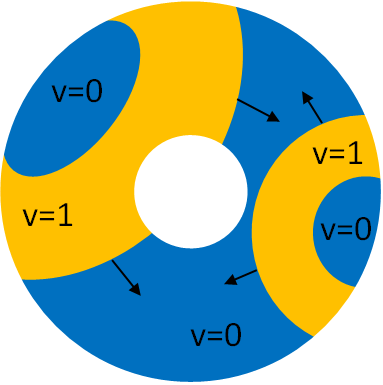}
\caption{The multi-source MEP}
\label{fig:propagation1}
\end{subfigure}
\begin{subfigure}{.23\textwidth}
\centering\includegraphics[width=1.4in]{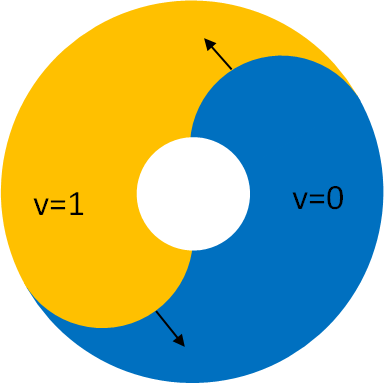}
\caption{The infinite MEP}
\label{fig:propagation2}
\end{subfigure}
\caption{Some possible MEP processes}
\label{fig:propagations}
\end{figure}

For an intuitive understanding of the basic problem, some kinds of MEP processes are shown in planar propagation spaces with holes in Fig.~\ref{fig:propagations}.
Firstly, Fig.~\ref{fig:propagation0} shows the simplest case where a well-initialized propagation is produced by an external trigger event occurring in some correct cell.
Fig.~\ref{fig:propagation1} shows the case where the external trigger events occur in more than one such cell.
Both cases are desired as the planar propagation spaces can be well partitioned into the desired propagation regions.

However, Fig.~\ref{fig:propagation2} shows some undesired cases where the propagation might not be able to terminate since the triggered cells can be restored before the trigger signals being propagated back to them again in the specific spaces.
In this case, the propagation cannot form an o-MEP since at least (\ref{eq:def_valid}) and (\ref{eq:def_simple}) are breached.
So, in the general cases, the system $\mathcal{S}$ may not be able to form the desired o-MEP, and there is no guarantee that the subsequent propagations would form the desired p-MEP process.
Thus, a practical problem is how to efficiently configure the time parameters of $\mathcal{S}$ to allow the desired p-MEP process to be formed with arbitrary initial system states in arbitrarily connected networks.
This is referred to as the self-stabilizing p-MEP problem.

Further, inspired by the reaction-diffusion processes and the classical work \cite{TURING1990153}, the MEP formed in $G$ may have various propagation patterns.
So, an additional problem is how the propagation patterns can evolve and be further controlled in the MEP systems.

\section{Self-stabilizing periodic mutual-exclusive propagation}
\label{sec:MEP}

Formally, we say a trigger event $u=\mathtt{e}_\mathtt{u}(i,t)$ occurs, denoted as $u\in \mathtt{E}_\mathtt{u}^i$, iff $t\in T_i$.
If a trigger signal is emitted by $i$ in responding to the occurrence of $u\in \mathtt{E}_\mathtt{u}^i$, we say the trigger signal is emitted by $i$ with $u$.
We say $u_1\in \mathtt{E}_\mathtt{u}^{i_1}$ reaches $u_2\in \mathtt{E}_\mathtt{u}^{i_2}$, denoted as $u_1\mapsto u_2$, iff the trigger signal emitted by $i_1$ with $u_1$ is accepted by $i_2\in N_{i_1}$ and $u_2$ is produced by this accepted trigger signal.
We say $u_1\in \mathtt{E}_\mathtt{u}^{i_1}$ is ended at $u_2\in \mathtt{E}_\mathtt{u}^{i_2}$, denoted as $u_1 \leadsto u_2$, iff the trigger signal emitted by $i_1$ with $u_1$ is rejected by $i_2\in N_{i_1}$ with $u_2$.
For convenience, the set of all trigger events occurring in the cell set $N$ is denoted as $\mathtt{E}_\mathtt{u}=\cup _{i\in N}\mathtt{E }_\mathtt{u}^i$.

\subsection{A finer constraint}
For self-stabilization of the MEP systems, we add the following constraint
\begin{eqnarray}
\label{eq:propa_0input} (1-\rho)\tau_1/3>\tau_0> (1+\rho)(L_G+1)d\\
\label{eq:propa_1input} \forall i\in N,\forall t_1\geqslant t_0:~~~~~~~~~~~~~~~~~~~~~~~~~~~~~~~~~~\nonumber\\
w(i,t_1)=1\to \forall t\in [t_1-\tau_1,t_1]:u(i,t )=0
\end{eqnarray}
where $L_G$ is the length of the longest simple path (without any loop) on $G$.

Now, suppose that (\ref{eq:sig_propa}), (\ref{eq:observe_j_i}), (\ref{eq:sig_propa_u1}), (\ref{eq:propa_0input}), and (\ref{eq:propa_1input}) are satisfied in $\mathcal{S}$.
We show that the desired p-MEP process can be formed in a bounded time with arbitrarily initial states.

For this, we first define the association relationship and strong association relationship of the trigger events on the undirected network $G$.
\begin{definition}
\label{def_propa_related}
For any $i_1,i_2\in N$, $u_1=\mathtt{e_u}(i_1,t_1)$ is associated with $u_2=\mathtt{e_u}(i_2,t_2)$ on $G$, denoted as $u_1\stackrel{G}{\sim} u_2$, iff any one the following conditions holds
\begin{eqnarray}
\label{def_propa_related1} u_1= u_2 \lor u_1\mapsto u_2 \lor u_1\leadsto u_2\\
\label{def_propa_related2} u_2\stackrel{G}{\sim} u_1 \lor \exists u_3\in \mathtt{E}_{\mathtt{u}}^{i}:i\in N \land u_1\stackrel{G}{\sim}u_3 \land u_3\stackrel{G}{\sim}u_2
\end{eqnarray}
\end{definition}
\begin{definition}
\label{def_propa_srelated}
For any $i_1,i_2\in N$, $u_1=\mathtt{e_u}(i_1,t_1)$ is strongly associated with $u_2=\mathtt{e_u}(i_2,t_2)$ on $G$, denoted as $u_1 \stackrel{G}{\simeq} u_2$, iff $u_1\stackrel{G}{\sim} u_2$ and any one of the following conditions holds
\begin{eqnarray}
\label{eq_propa_srelated1} (i_1,i_2)\in E\land |t_1-t_2|\leqslant d\\
\label{eq_propa_srelated2} \exists u_3\in \mathtt{E}_{\mathtt{u}}^{i}:i\in N \land u_1\stackrel{G}{\simeq}u_3 \land u_3\stackrel{G}{\simeq}u_2
\end{eqnarray}
\end{definition}
For any two trigger events $u_1,u_2$, we use $u_1\stackrel{G}{\nsim} u_2$ to indicate that $u_1\stackrel{G}{\simeq} u_2$ does not hold.
When it is not confused, $\stackrel{G}{\sim}$ and $\stackrel{G}{\simeq}$ are simplified as $\sim$ and $\simeq$, respectively.
Obviously, for any undirected graph $G=(N,E)$, $\sim$ and $\simeq$ are equivalent relations.
Meanwhile, $\sim$ and $\simeq$ also have the following basic properties.
\begin{lemma}
\label{lemma_propa_related_all}
If $G=(N,E)$ is connected, then $\forall u\in \mathtt{E}_{\mathtt{u}}:\forall i\in N:\exists u'\in \mathtt{E}_{\mathtt{u}}^{i}:u\stackrel{G}{\sim}u'$.
\end{lemma}
\begin{IEEEproof}
Let $u=\mathtt{e_u}(i_0,t_0)$.
By definition we have $u\stackrel{G}{\sim}\mathtt{e_u}(i_0,t_0)$.
For $\forall i_1\in N_{i_0}$, the trigger signal emitted by $i_0$ with $u$ is either accepted or rejected by $i_1$.
If this trigger signal is accepted by $i_1$, there exists $u_1'\in \mathtt{E}_{\mathtt{u}}^{i_1}$ satisfying $u\mapsto u_1'$.
Otherwise, if this trigger signal is rejected by $i_1$, with (\ref{eq:sig_propa}) there is $u_1''\in \mathtt{E}_{\mathtt{u}}^{i_1 }$ satisying $u\leadsto u_1''$.
So $\exists u_1\in \mathtt{E}_{\mathtt{u}}^{i_1}:u\stackrel{G}{\sim}u_1$ holds in all cases.
Similarly, for $\forall i_2\in N_{i_2}$, we have $\exists u_2\in \mathtt{E}_{\mathtt{u}}^{i_2}:u_1\stackrel{G}{ \sim}u_2$.
As $G$ is a connected, we have $\forall i\in N:\exists t:u\stackrel{G}{\sim}\mathtt{e_u}(i,t)$.
\end{IEEEproof}

\begin{lemma}
\label{lemma_propa_srelated_partition}
For any two cells $i_1,i_2\in N$ on $G=(N,E)$, if there are $u_1=\mathtt{e_u}(i_1,t_1)$ and $u_2= \mathtt{e_u}(i_2,t_2)$ satisfying $u_1 \stackrel{G}{\nsim} u_2$ and $u_1 \stackrel{G}{\sim} u_2$, then there exist $E'\subseteq E$, $1<m\leqslant n$, $1\leqslant m_1\ne m_2\leqslant m$ such that $G-E'$ is composed of two or more connected components (or saying the nonempty maximal connected subgraphs) $G_1=(N_1,E_1),\dots,G_m=(N_m,E_m)$, $i_1\in N_{m_1}$, $i_2\in N_{m_2}$, and $\forall (i_1',i_2')\in E':(\mathtt{e_u}(i_1',t_1')\stackrel{G} {\sim}u_1 \land \mathtt{e_u}(i_2',t_2')\stackrel{G}{\sim}u_2)\to |t_1'-t_2'|>d$.
\end{lemma}
\begin{IEEEproof}
Suppose there is no such edge set $E'$.
As $G$ is connected, there exist a simple path $(i_1',i_2',\dots,i_s')$ from $i_1'=i_1$ to $i_s'=i_2$ on $G$ and $u_1=\mathtt{e_u}(i_1',t_1')\sim\mathtt{e_u}(i_2',t_2')\sim\dots\sim\mathtt{e_u}(i_s',t_s')=u_2$ satisfying $\forall k\in \{1,2,\dots,s-1\}:|t_k'-t_{k+1}'|\leqslant d$.
So we have $u_1\stackrel{G}{\simeq} u_2$, which contradicts $u_1 \stackrel{G}{\nsim} u_2$.
\end{IEEEproof}

Now we show that the associated trigger events occur at some near instants.
Intuitively, for the two-cells undirected propagation network $K_2=(\{i_1,i_2\},\{(i_1,i_2)\})$, we have the following property.
\begin{lemma}
\label{lemma_propa_related_near2}
$(\mathtt{e_u}(i_1,t_1)\stackrel{K_2}{\sim}\mathtt{e_u}(i_2,t_2))\to|t_1-t_2|\leqslant d$
\end{lemma}
\begin{IEEEproof}
Suppose that $u_1=\mathtt{e_u}(i_1,t_1)$, $u_2=\mathtt{e_u}(i_2,t_2)$, and $u_1\mapsto u_2 \lor u_1\leadsto u_2$ holds (the case $ u_2\mapsto u_1 \lor u_2\leadsto u_1$ can be handled similarly).
If $u_1\mapsto u_2$, the conclusion holds obviously.
If $u_1\leadsto u_2$, we have $t_1-t_2\in [0,\tau_0/(1-\rho)]$.
In this case, if $t_1-t_2\in (d,\tau_0/(1-\rho)]$, the trigger signal emitted by $i_2$ with $u_2$ would be observed by $i_1$ at some instant $t_3\in [t_2,t_2+d]$.
So $i_1$ either accepts or rejects this trigger signal at $t_3$.
If $i_1$ accepts the trigger signal, then $u_3=\mathtt{e_u}(i_1,t_3)$ occurs at $t_3$.
If $i_1$ rejects the trigger signal, then $u_4=\mathtt{e_u}(i_1,t_4)$ occurs at some instant $t_4\in [t_3-\tau_0/(1-\rho),t_3 ]$.
As $t_3,t_4\in [t_1-2\tau_0/(1-\rho),t_1]$, with $\tau_1>3\tau_0/(1- \rho)$ required in (\ref{eq:propa_0input}) (and (\ref{eq:propa_1input})), $u_1$ is not produced by any $w$ event.
Now as $i_1$ has the unique neighbor $i_2$, there should be some $u_5=\mathtt{e_u}(i_2,t_5)$ satisfying $u_5\mapsto u_1$ and $t_5\in [t_1-d,t_1]$.
Meanwhile, as $t_2\in [t_1-\tau_0/(1-\rho),t_1-d)$, again with (\ref{eq:propa_0input}), $u_5$ is not produced by any $w$ event.
Similarly, as $i_2$ has the unique neighbor $i_1$, there should be $u_6=\mathtt{e_u}(i_1,t_6)$ satisfying $u_6\mapsto u_5$ and $t_6\in [t_5-d,t_5]$.
Thus, both $u_1$ and $u_6$ occur in $i_1$ with $t_1-t_6\in [0,2d]$, and $u_1$ is not produced by any $w$ event.
This contradicts $\tau_0>2d (1+\rho)$ required in (\ref{eq:propa_0input}).
So we have $|t_1-t_2|\leqslant d$.
\end{IEEEproof}

Now we show that a similar result holds for any undirected network $G$.
\begin{lemma}
\label{lemma_propa_related_near}
$\mathtt{e_u}(i_1,t_1)\stackrel{G}{\sim}\mathtt{e_u}(i_2,t_2)$ holds iff $\mathtt{e_u}(i_1,t_1)\stackrel{G}{\simeq}\mathtt{e_u}(i_2,t_2)$.
\end{lemma}
\begin{IEEEproof}
Denote $u_1=\mathtt{e_u}(i_1,t_1)$, $u_2=\mathtt{e_u}(i_2,t_2)$, and $U=\{u\in \mathtt{E}_{\mathtt{u}}\mid u\stackrel{G}{\sim}u_1\}$.
By definition $(u_1\stackrel{G}{\simeq}u_2)\to (u_1\stackrel{G}{\sim}u_2)$ holds.
Thus, we assume $u_1\stackrel{G}{\sim}u_2$ and need only to show $u_1\stackrel{G}{\simeq} u_2$.
Suppose that $u_1\stackrel{G}{\simeq} u_2$ does not hold in this case.
With Lemma \ref{lemma_propa_srelated_partition}, there exists $E_1'\subseteq E$ satisfying $G-E_1'$ being with two or more connected components $G_1'=(N_1,E_1),\dots ,G_m'=(N_m,E_m)$ and $\forall (i_1',i_2') \in E_1':\forall \mathtt{e_u}(i_1',t_1'),\mathtt{e_u}(i_2',t_2')\in U: |t_1'-t_2'|>d$.
With Lemma \ref{lemma_propa_related_all}, there exist $(j_1',j_2') \in E_1'$ and $u_1'=\mathtt{e_u}(j_1',t_1'),u_2'=\mathtt{e_u}(j_2',t_2')\in U$ satisfying $u_1'\leadsto u_2'$ or $u_2'\leadsto u_1'$, for otherwise $(\mathtt{e_u}(i_1,t_1)\stackrel{G}{\sim}\mathtt{e_u}(i_2,t_2))$ cannot hold.
Without loss of generality, suppose that $u_1'\leadsto u_2'$ (the case $u_2'\leadsto u_1'$ can be handled similarly), $j_1'\in N_1$, and $G_1=G_1'=(N_1,E_1)$.
Then, as $|t_1'-t_2'|>d$, we have $t_2'\in [t_1'-\tau_0/(1-\rho),t_1'-d)$.
Therefore, when the trigger signal emitted by $j_2'$ with $u_2'$ is observed by $j_1'$ in $[t_1'-\tau_0/(1-\rho),t_1')$, with (\ref{eq:sig_propa}), $j_1'$ cannot be triggered in $[t_1'-\tau_0/(1 +\rho),t_1')$, since $j_1'$ is triggered at $t_1'$.
So $j_1'$ can only be triggered at some $t_1'''\in[t_1'-2\tau_0/(1-\rho),t_1'-\tau_0/(1+\rho))$.
Denote $u_1''= \mathtt{e_u}(j_1',t_1''')$.
With $u_1''\ne u_1'$ and $\tau_1>3\tau_0/(1-\rho)$ required in (\ref{eq:propa_0input}), $u_1'$ is not produced by any $w$ event.
Meanwhile, the trigger signals emitted by the cells outside $G_1$ with the trigger events in $U$ cannot be accepted by the cell in $G_1$, for otherwise there is $(j_1'',j_2'')\in E_1'$ satisfying $\mathtt{e_u}(j_1'',t_1'')\stackrel{G} {\simeq}\mathtt{e_u}(j_2'',t_2'')$ in $U$.
So, there is a cell $i_0$ in $G_1$ satisfying $i_0\ne j_1'$, $s(\gamma_{j_1'}(t_1'))=i_0$, and the path $P''=\gamma_{j_1'}(t_1')=(j_0''=i_0,j_1'',j_2'',\dots,j_r''=j_1')$ is in $G_1$.
Denoting $u_0=\mathtt{e_u}(i_0,t_0)$ as the source trigger event in forming the propagation path $P''$, we have $u_0\stackrel{G_1}{\sim} u_1' \stackrel{G}{\sim} u_2'\stackrel{G}{\sim} u_1''$.
Thus we have $u_0\stackrel{G}{\sim} u_1''$.
Meanwhile, as $u_1''$ occurs, there exists some $u_0'=\mathtt{e_u}(i_0,t_0')$ satisfying $u_1''\stackrel{G_1}{\sim} u_0'$.

Now we show that $u_1''\stackrel{G_1}{\simeq} u_0'$ cannot hold for any possible $u_0'$.
Firstly, for $u_0'=u_0$, as $u_0\stackrel{G_1}{\simeq} u_1'$ and $u_1''\ne u_1'$, with $L_{G_1}d<\tau_0/(1+\rho)$ required in (\ref{eq:propa_0input}), $u_1''\stackrel{G_1}{\simeq} u_0'$ cannot hold.
And for $u_0'\ne u_0$, suppose that $u_1''\stackrel{G_1}{\simeq} u_0'$.
Then there is a simple path $P=(j_0=j_1',j_1,j_2,\dots,j_s=i_0)$ in $G_1$ making $u_1''\stackrel{G_1}{\simeq} u_0'$ with satisfying $(j_k,j_{k+1})\in E_1\land |t_1-t_2|\leqslant d$ required in (\ref{eq_propa_srelated1}).
As a $w$ event occurs in $i_0$ at $t_0$, with $\tau_1>3\tau_0/ (1-\rho)$ required in (\ref{eq:propa_0input}), it can only be $t_0'>t_0$ and thus $t_0'-t_0> (L_G+1)d$.
Meanwhile, as $u_1'\ne u_1''$ and $t_1'>t_1''$, we also have $t_1'-t_1''> (L_G+1)d$.
This cannot hold together with $|t_0-t_1'|\leqslant (L_G+1)d$ and $|t_0'-t_1''|\leqslant (L_G+1)d$.

Thus, there exist $N_1\subset N$, $i_0,j_1'\in N_1$, $i_0\ne j_1'$, $u_{10}=\mathtt{e_u}(i_0,t_0')$ and $u_{11}= \mathtt{e_u}(j_1',t_1''')$ satisfying $u_{11}\stackrel{G_1}{\nsim}u_{10}\land u_{11}\stackrel{G_1}{\sim}u_{10}$.
Meanwhile, the trigger signals emitted by the cells outside $G_1$ with the trigger events in $U$ cannot be accepted by the cell in $G_1$.
So, by repeatedly applying the above steps, there exist a nonempty connected graph $G_2=(N_2,E_2)$, $N_2\subset N_1$, such that some trigger events $u_{21},u_{20}$ satisfy $ u_{21}\stackrel{G_2}{\nsim}u_{20}\land u_{21}\stackrel{G_2}{\sim}u_{20}$, and the trigger signals emitted by the cells outside $G_2$ with the trigger events in $U$ cannot be accepted by the cell in $G_2$.
Iteratively, there are nonempty connected graphs $G_h=(N_h,E_h)$, $N_h\subset N_{h-1}\subset\dots\subset N_2\subset N_1$, which have similar properties.
As $|N|$ is finite, there is $h'<|N|$ making such $G_{h'}=(N_{h'},E_{h'})$ and $|N_{h'}|\leqslant 2$.
If $|N_{h'}|=2$, with Lemma \ref{lemma_propa_related_near2}, we have $u_{h'1}\stackrel{G_1}{\simeq}u_{h'0}$.
This contradicts $u_{h'1}\stackrel{G_1}{\nsim}u_{h'0}$.
And if $|N_{h'}|=1$, there cannot be $i,j\in N_{h'}$ and $i\ne j$, which contradicts the basic properties of $G_{h'}$.
Thus $u_1\stackrel{G}{\nsim} u_2$ cannot hold.
\end{IEEEproof}

\begin{corollary}
\label{corollary_propa_related_near}
$(\mathtt{e_u}(i_1,t_1)\stackrel{G}{\sim}\mathtt{e_u}(i_2,t_2))\to |t_1-t_2|\leqslant L_G d$
\end{corollary}
\begin{IEEEproof}
With $\mathtt{e_u}(i_1,t_1)\stackrel{G}{\sim}\mathtt{e_u}(i_2,t_2)$ and Lemma~\ref{lemma_propa_related_near}, we have $\mathtt{e_u}(i_1,t_1)\stackrel{G}{\simeq}\mathtt{e_u}(i_2,t_2)$.
Thus, by definition we have $|t_1-t_2|\leqslant L_G d$.
\end{IEEEproof}

It should be noted that the above results are established with (\ref{eq:propa_0input}), in which the external trigger time parameter $\tau_1$ can be approximately configured as $3$ times the local restoration threshold parameter $\tau_0$ (with the bounded drift rates being ignored).
In \cite{YuCOTS2021}, a coarser result is established by mainly configuring a sufficiently long user stage (the TT stage), as the user stage can be much longer than $\tau_0$ in building higher-layer TT communication with MEP.
Here the result is provided in a self-contained way.
We can see that, in considering the worst cases, the stabilization time of $\mathcal{S}$ mainly depends on $L_G$.
In the worst case, we have $L_G=n-1$ when $G$ is a Hamilton graph.
But in some other cases, for example, $G$ being some $r$-ary trees, $L_G$ can be within $O(\log n)$.
\subsection{Self-stabilizing periodic MEP process}

For reaching the desired periodic MEP process, (\ref{eq:propa_separation}), (\ref{eq:I_N}), (\ref{eq:propa_separation_mutex}), and (\ref{eq:propa_separation_mutex_liveness}) should be satisfied with sufficiently good precision.
Meanwhile, for reaching fast self-stabilization, (\ref{eq:sig_propa}), (\ref{eq:observe_j_i}), (\ref{eq:propa_0input}), and (\ref{eq:propa_1input}) should also be satisfied as soon as possible with arbitrary initial system states.
To this, firstly, the local clock in each cell can be employed to measure the elapsed local time (in local ticks) since the last trigger event occurred in this cell.
Then, when the elapsed local time is equal to or greater than some local liveness threshold, for example $\tau_2=\tau_1(1+\rho)$, the cell should be externally triggered.
As the local liveness threshold $\tau_2$ is bounded, when the measured time is out of the valid range (for example, greater than $\tau_2$), the cell can be recovered from invalid time-measurement immediately.
Thus, a timer configured with a constant local duration $\tau$ can always be recovered within $\tau/(1-\rho)$ time.
In other words, we can make self-stabilizing timers from the local clocks with bounded drift rates.
In this sense, we say a timer of cell $i$ is stabilized if it correctly measures the elapsed local time since the last trigger event occurred in $i$.
With the stabilized timers, the nonfaulty cells are expected to realize the given propagation constraints.
Similar to \cite{YuCOTS2021}, we say a cell $i\in N$ is correct since $t$ iff $i$ is nonfaulty and the timers of $i$ are stabilized since $t$.
As the local restoration threshold $\tau_0$ is also measured since the last trigger event and $\tau_0<\tau_2$ holds, a nonfaulty cell can become correct within $\tau_2/(1-\rho)$.
With this, we show that $\mathcal{S}$ can always form the sp-MEP process with bounded errors in a bounded time.
Without loss of generality, we assume the MEP system $\mathcal{S}$ runs with an arbitrary initial state since $t_0=0$.

\begin{lemma}
\label{lemma_propa_mutex_separation_cycle}
There exists some $t\in [t_2,t_2+\tau_2]$, $t_2=(\tau_2+\tau_0)/(1-\rho)+d+\tau_1$, that $\mathcal{S}$ forms a $(\tau_\pi,\tau_\Delta,\tau_\nabla)$ p-MEP process since $t$, where $\tau_\pi\leqslant D_G d$, $\tau_\Delta=\tau_1$, $\tau_\nabla=\tau_2/(1-\rho)$, and $D_G$ is the diameter of $G$.
\end{lemma}
\begin{IEEEproof}
Firstly, although the states of the timers in $\mathcal{S}$ are arbitrary at instant $0$, the cells can only be triggered by stabilized timers since $t_0=\tau_2/(1-\rho)+d$.
Thus, since $t_1=t_0+\tau_0/(1-\rho)$, all new $u$ and $w$ events would satisfy (\ref{eq:sig_propa}) and (\ref{eq:propa_1input}).
So, since $t_2=t_1+\tau_1$, the propagation process in $\mathcal{S}$ would be restricted by the results of Lemma~\ref{lemma_propa_related_near} and Corollary~\ref{corollary_propa_related_near}, as the time referenced in the proofs of Lemma~\ref{lemma_propa_related_near} and Corollary~\ref{corollary_propa_related_near} is no earlier than $t_2-\tau_1$.
Namely, since $t_2$, if any trigger event $u$ occurs, all trigger events associated with $u$ would occur within $L_G d$ duration.
As $L_G d<\tau_0/(1+\rho)$, each cell generates at most $1$ trigger event in this duration.
Meanwhile, with Lemma~\ref{lemma_propa_related_all}, each cell generates at least $1$ trigger event in this duration.
Thus, all trigger events associated with $u$ would form a propagation satisfying (\ref{eq:propa_separation_mutex}) with some $\tau_\pi$ and $\tau_\Delta=\tau_1$.
Then, with (\ref{eq:sig_propa}), the next $u$ event occurring in every cell would no later than $\tau_\nabla=\tau_2/(1-\rho)$ since the last $u$ event in the same cell.
And when the next $u$ event occurs, as $\tau_2= \tau_1(1+\rho)$, we have $\forall i\in N:v(i,t^-)=1$.
As $\tau_0>(1+\rho)(L_G+1)d\geqslant 2(1+\rho)d$, a new o-MEP is formed.
Meanwhile, as $\forall i\in N:v(i,t^-)=1$, there is always $t_i'\leqslant t_j'\leqslant t_i'+d$ or $t_j'\leqslant t_i'\leqslant t_j'+d$ for any pair of neighbor cells $(i,j)$ with $\mathtt{e_u}(i,t_i'){\sim}\mathtt{e_u}(j,t_j')$.
So there is $|t_i'- t_j'|\leqslant d$ and thus $\tau_\pi\leqslant D_G d$.
\end{IEEEproof}

Intuitively, just as a combustion engine with multiple strokes can generate periodic power, the MEP system $\mathcal{S}$ can also generate periodic trigger events in the desired self-stabilizing way.

\section{Propagation patterns and local algorithms}
\label{sec:Pattern}
In the definition of o-MEP, the propagation paths can be arbitrarily formed in satisfying (\ref{eq:def_valid}) to (\ref{eq:def_propagative}).
Sometimes, better propagation paths can be formed in the MEP systems.
For example, when all cells are synchronized with sufficiently good precision, the propagation paths are expected to be significantly shorter than $D_G$.
Here, we first discuss how the propagation patterns of the p-MEP processes can naturally evolve without any additional controls.
Then, we discuss how the propagation patterns can be further controlled with some simple local algorithms.

\subsection{Propagation patterns}
The formation of some specific animal skin patterns is intuitively investigated with the propagation of chemical substances and reactions in the pioneering work \cite{TURING1990153} with the discrete reaction-diffusion processes.
Since (\ref{eq:sig_propa}) also describes a very simple kind of discrete reaction-diffusion processes, here it is interesting to investigate the formation of MEP patterns.
This investigation would have benefits for the applications of the MEP systems.

To describe the propagation pattern of an o-MEP $\mathcal{P}\in \mathbb{M}_G$, we divide the neighbor cells $j\in N_i$ of each cell $i$ into the following four categories (with respect to $\mathcal{P}$, the same below).

\begin{enumerate}[N1)]
\item
$j$ is the parent neighbor of $i$ iff $h_i=j$;
\item
$j$ is a child neighbor of $i$ iff $h_j=i$;
\item
$j$ is the alien neighbor of $i$ iff $s(\gamma_i)\ne s(\gamma_j)$;
\item
$j$ is the family neighbor of $i$ iff $j$ is in none of the above categories.
\end{enumerate}

It is easy to verify that any $j\in N_i$ belongs to one and only one of the above four kinds of neighbors of $i$.
With this, we can define the following four types of cells in $\mathcal{P}$ with the parent and child neighbors.

\begin{enumerate}[C1)]
\item
$i$ is a source cell, denoted as $i\in N_{\text{source}}(\mathcal{P})$, iff $i$ has no parent neighbors;
\item
$i$ is a sink cell, denoted as $i\in N_{\text{sink}}(\mathcal{P})$, iff $i$ has no child neighbors;
\item
$i$ is a flow cell, denoted as $i\in N_{\text{flow}}(\mathcal{P})$, iff $i$ has parent and child neighbors;
\item
$i$ is a united cell, denoted as $i\in N_{\text{united}}(\mathcal{P})$, iff $i$ has no parent nor child neighbors.
\end{enumerate}

In the o-MEP $\mathcal{P}$, obviously we have $N_{\text{united}}(\mathcal{P})=N_{\text{source}}(\mathcal{P}) \cap N_{\text{sink}}(\mathcal{P})$ and $N_{\text{flow}}(\mathcal{P})=N\setminus (N_{\text{source}} (\mathcal{P}) \cup N_{\text{sink}}(\mathcal{P}))$.
In addition, we can also differentiate the cells with family and alien neighbors:

\begin{enumerate}[C'1)]
\item
$i$ is a bank cell, denoted as $i\in N_{\text{bank}}(\mathcal{P}) $, iff $i$ has foreign neighbors;
\item
$i$ is a ridge cell, denoted as $i\in N_{\text{ridge}}(\mathcal {P})$, iff $i$ has family but no foreign neighbors;
\item
$i$ is a flat cell, denoted as $i\in N_{\text{flat}}(\mathcal {P})$, iff $i$ has no foreign nor family neighbors.
\end{enumerate}

Obviously, in the o-MEP $\mathcal{P}$, any cell $i\in N$ belongs to one and only one of the above three kinds of cells.
For convenience, when it is not confused, $N_{\mathtt{x}}(\mathcal{P})$ is simplified as $N_{\mathtt{x}}$.
Moreover, we use $N_{\mathtt{X}}=\cup _{\mathtt{x}\in \mathtt {X}}N_{\mathtt{x}}$ to represent the generalized $\mathtt{X}$ pattern of $\mathcal{P}$.
For example, given an o-MEP $\mathcal{P}$, the \emph{bank ridge} pattern of $\mathcal{P}$ is $N_{\text{bank}}\cup N_{\text{ridge}}$, the \emph{sink flow} pattern of $\mathcal{P}$ is $N_{\text{sink}}\cup N_{\text{flow}}$, and the \emph{non-source} pattern of $\mathcal{P}$ is $N\setminus N_{\text{source}}$, etc..

\begin{lemma}
\label{lemma_propa_mutex_cells}
The patterns of an o-MEP $\mathcal{P}$ have the following basic properties.
\begin{enumerate}[BP1)]
\item
There is at least one sink cell in each propagation region.
\item
The sink cells are not less than the source cells.
\item
If there are non-sink (not sinking) cells in a propagation region, then there exist non-sink source cells in this propagation region.
\item
If $G$ is a tree, then there is no ridge cell.
\item
A $d$-degree (or saying \emph{valence}) flat cell with $d\geqslant 2$ is not a sink cell.
\item
If a source cell has $m$ child neighbors, the propagation region where the source cell is located has at least $m$ sink cells.
\item
If there are $m$ $d$-degree ($d\geqslant 2$) flat cells, then there are at least $(d-2)m+1$ sink cells.
\end{enumerate}
\end{lemma}
\begin{IEEEproof}
If there is no sink cell in a propagation region, there would be a propagation loop of the trigger signals, which contradicts the definition of the o-MEP.
And since there is a unique source cell in each propagation region, the sink cells are not less than the source cells.
If all non-sink cells existing in a propagation region are also non-source, then there also be a propagation loop of the trigger signals.
If there is a ridge cell on tree $G$, there is a mutually rejected cell pair $(i,i')$ such that $s(\gamma_i)=s(\gamma_{i'})$, and thus a ring exists in $G$.
If the $2$-degree (and above) flat cell $i$ is the sink cell, then $i$ has $d$ different parent neighbors, which contradicts the definition of the o-MEP.
Since the propagation in each propagation region can form a directional propagation tree based on the parent-child relationship of the cells, the leaf cells are no less than the children of the tree root, with which the propagation region with an $m$-degree source cell has at least $m$ sink cells.
As a $d$-degree flat cell with $d\geqslant 2$ has $d-1$ child neighbors, and the source cell has at least one child neighbor (for otherwise there is no flat cell) in the same directed propagation tree, there are at least $(d-2)m+1$ sink cells in this tree.
\end{IEEEproof}

\subsection{Natural formations of propagation patterns}

As is shown, $(\tau_\pi,\tau_\Delta,\tau_\nabla)$ p-MEP processes can be formed in properly configured $\mathcal{S}$ with bounded message delays and bounded clock drift rates.
Considering some real-world applications, the precision $\tau_\pi$ is expected to be as small as possible.
Obviously, if the actual signal delays can be precisely estimated in all cells, $\tau_\pi$ can be improved with globally broadcasted report messages \citep{YuCOTS2021}.
However, without relying on precise delay measurements and global message broadcasts, the propagation process can only be locally observed with the trigger signals in each cell.
In some situations, we might expect the most favorable results with the least cost.
Practically, this expectation of the MEP systems largely depends on the tempers of the \emph{natural} propagations.
To utilize the properties of the so-called \emph{natural} propagations, we investigate the self-contained MEP system $\mathcal{S}_0$ with the signal delays being randomly distributed in $[d_{min},d_{max}]$ with $0\leqslant d_{min}\leqslant d_{max}=d$.
We would see that, although the worst-case precision $\tau_\pi$ would be at the order of $D_G d$, one can often expect much better with the randomly distributed signal delays.

For the stabilized $\mathcal{S}_0$, denote the $k$th o-MEP of the sp-MEP process (since stabilization) as $\mathcal{P}(k)=\{\mathtt{e_u}(i,t_{k}^{(i)})\mid i\in N\}$ with the trigger instant $t_{k}^{(i)}$ for each $i\in N$.
Denote $T(\mathcal{P}(k))=\{t_{k}^{(i)}\mid i\in N\}$ and $t_{k}^{(min)}=\min\{T(\mathcal{P}(k))\}$,
Then, by defining the propagation error of $\mathcal{P}(k)$ as $e_1(k)=\sum_{t\in T(\mathcal{P}(k))}(t-t_{k}^{(min)})$ and ignoring the bounded clock drifts, as only the timers of the internally triggered cells would be adjusted towards their pioneers, we have $e_1 (k+1)\leqslant e_1(k)$ for any $k\geqslant 1$.
Namely, if the bounded drifts of the clocks can be ignored, $e_1$ would monotonically decrease with the increase of $k$ since the stabilization of the sp-MEP system.
Further, for a bounded convergence time, it is highly desired that $e_1(k+1)\leqslant e_1(k)-a$ holds with a sufficiently large $a>0$ when $e_k$ is large.

Concretely, with ignoring the bounded clock drifts, for $e_1(k+1)< e_1(k)$, there should be at least one non-source cell (being internally triggered) in $\mathcal{P}(k+1)$.
Obviously, when the delays are controlled by a malicious adversary who can arbitrarily control the actual signal delays in $[0,d]$, there is no guarantee that $e_1(k+1)$ would be further reduced when $e_1(k)\leqslant D_G d$.
However, when the signal delays are randomly distributed, the situation can be much different.
To show this, we run some numeric simulations of $\mathcal{S}_0$ with $G$ being networks of some specific topologies.
In all simulations, with $d=10^{-3}~s$, the related timer parameters are configured approximately as $\tau_0=(1+\rho)(L_G+2)d$, $\tau_2=3(1+\rho)(\tau_0/(1-\rho)+d)$ with some specific $\rho$.
The initially elapsed local time recorded in each timer is randomly configured with the independent uniform distribution $U[0,\tau_2]$.

\begin{figure}[htbp]
\centering
\begin{subfigure}{.23\textwidth}
\centering\includegraphics[width=1.8in]{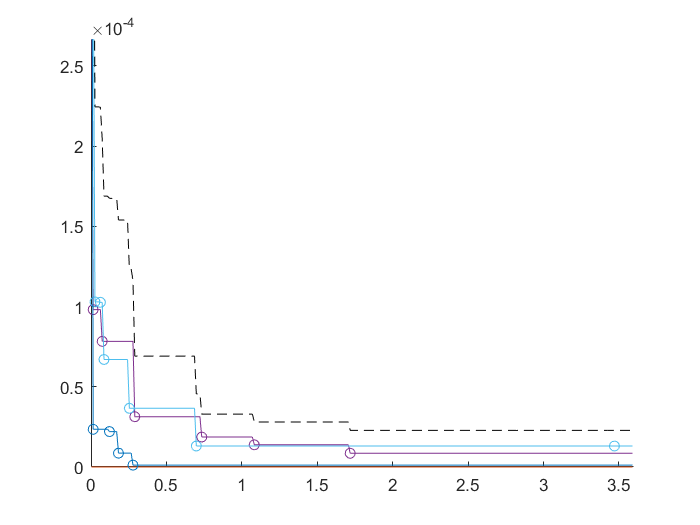}
\caption{for a $4$-cell ring network}
\label{fig:sim_natural_ring0}
\end{subfigure}
\begin{subfigure}{.23\textwidth}
\centering\includegraphics[width=1.8in]{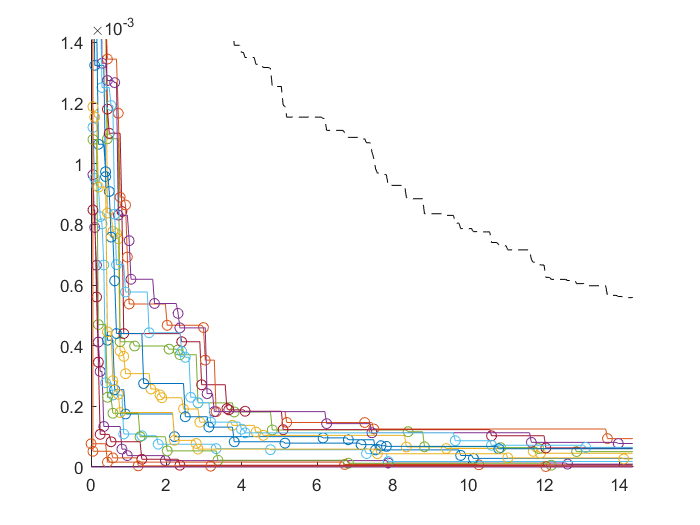}
\caption{for a $16$-cell ring network}
\label{fig:sim_natural_ring1}
\end{subfigure}
\begin{subfigure}{.23\textwidth}
\centering\includegraphics[width=1.8in]{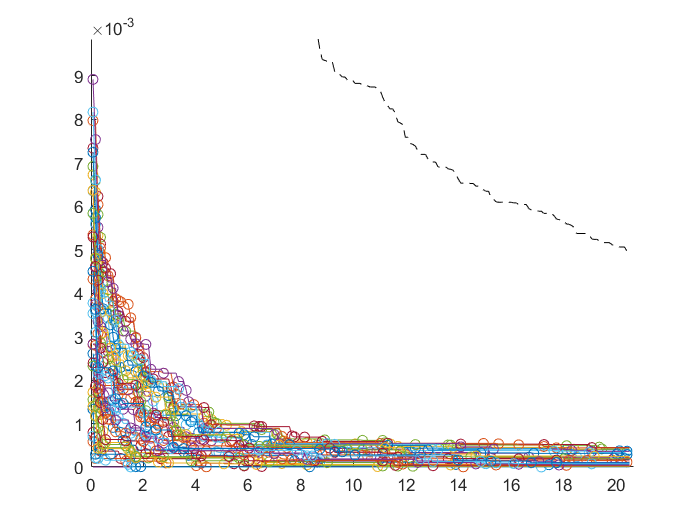}
\caption{for a $32$-cell ring network}
\label{fig:sim_natural_ring2}
\end{subfigure}
\begin{subfigure}{.23\textwidth}
\centering\includegraphics[width=1.8in]{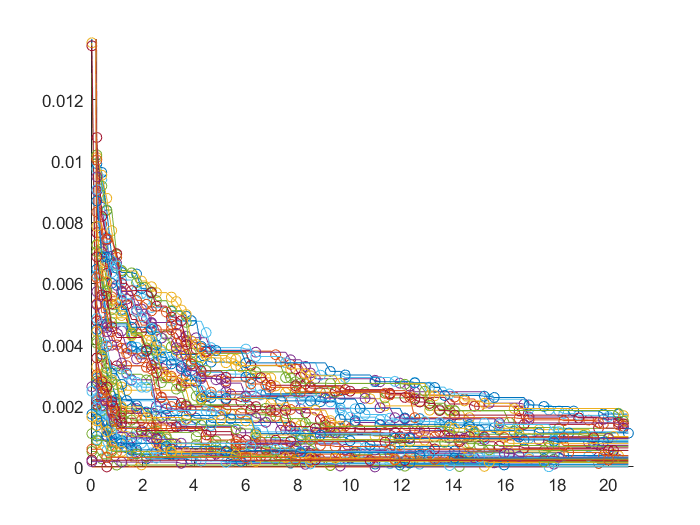}
\caption{for a $64$-cell ring network}
\label{fig:sim_natural_ring3}
\end{subfigure}
\caption{The sp-MEP processes with ring topologies}
\label{fig:sim_natural_ring}
\end{figure}

Firstly, in Fig.~\ref{fig:sim_natural_ring0}, Fig.~\ref{fig:sim_natural_ring1}, Fig.~\ref{fig:sim_natural_ring2}, and Fig.~\ref{fig:sim_natural_ring3}, $\mathcal{S}_0$ is simulated with the signal delays being independently sampled with the uniform distribution $U[0,d]$, $d=10^{-3}~s$, $\rho=0$, and $G$ being the ring topologies with $4$,  $16$, $32$, and $64$ cells, respectively.
It shows that, firstly, despite the random initial states of the system $\mathcal{S}_0$, $(\tau_\pi,\tau_\Delta,\tau_\nabla)$ p-MEP processes can be formed in some time since the startup of the system with corresponding to the result of Lemma~\ref{lemma_propa_mutex_separation_cycle}.
With this, to show the changes of the propagation errors $e_1(k)$ in the p-MEP process with the progress of $k$, we choose the earliest trigger instant $t_{k}^{(min)}$ in $\mathcal{P}(k)$ for each $k\geqslant 1$ as the values taken in the $x$-axis and show $\tilde t_{k}^{(i)}= t_{k}^{(i)}-t_{k}^{(min)}$ for every $i\in N$ in the $y$-axis.
Meanwhile, the propagation errors $e_1(k)=\sum_{i\in N}\tilde t_{k}^{(i)}$ are also calculated and represented as the dotted curves.
Thus, in all curves shown in Fig.~\ref{fig:sim_natural_ring}, the leftmost points show the results of the $(k=1)$st o-MEP of the sp-MEP processes, and the rightmost ones show the results with some larger $k$.
By representing the points of non-source cells in every o-MEP as circles, we see that the reduction of $e_1(k)$ results from the internal trigger events occurring in the non-source cells.
With this, the errors $\tilde t_{k}^{(i)}$ for all $i\in N$ can be reduced to a small fraction of $d$ with not much large $k$ in the $4$-cell, $16$-cell, and $32$-cell ring networks.
However, with the increase of $n$, the convergence of the system $\mathcal{S}_0$ with the ring topologies tend to be slow.
Nevertheless, the errors $\tilde t_{k}^{(i)}$ for all $i\in N$ can be fast reduced to the range of several $d$ even in the $64$-cell ring network.

\begin{figure}[htbp]
\centering
\begin{subfigure}{.23\textwidth}
\centering\includegraphics[width=1.8in]{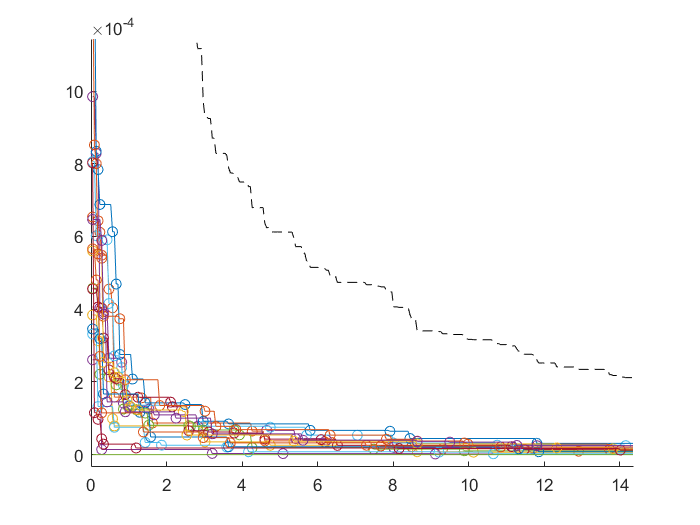}
\caption{for a $16$-cell gird network}
\label{fig:sim_natural_gird1}
\end{subfigure}
\begin{subfigure}{.23\textwidth}
\centering\includegraphics[width=1.8in]{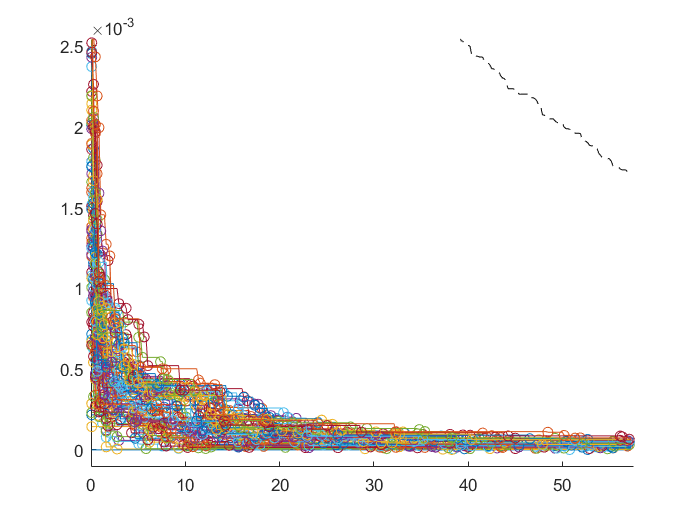}
\caption{for a $64$-cell gird network}
\label{fig:sim_natural_gird2}
\end{subfigure}
\begin{subfigure}{.23\textwidth}
\centering\includegraphics[width=1.8in]{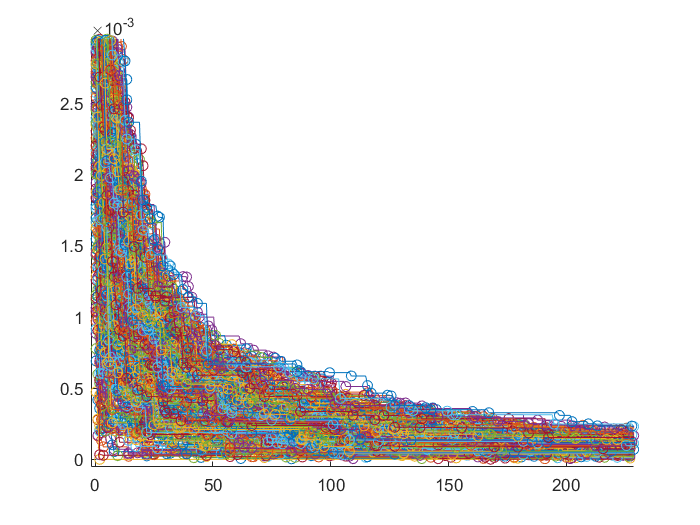}
\caption{for a $256$-cell gird network}
\label{fig:sim_natural_gird3}
\end{subfigure}
\begin{subfigure}{.23\textwidth}
\centering\includegraphics[width=1.8in]{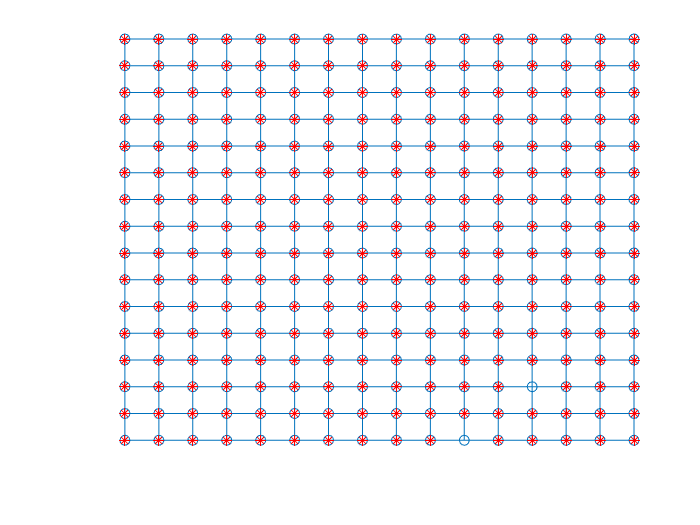}
\caption{the \emph{source} pattern}
\label{fig:sim_natural_gird3_pattern}
\end{subfigure}
\caption{The sp-MEP processes with gird topologies}
\label{fig:sim_natural_gird}
\end{figure}

For faster convergence, in Fig.~\ref{fig:sim_natural_gird1}, Fig.~\ref{fig:sim_natural_gird2}, and Fig.~\ref{fig:sim_natural_gird3}, we run the simulation with the same system settings as in Fig.~\ref{fig:sim_natural_ring} except for $G$ being networks of gird topologies with $16$, $64$, and $256$ cells, respectively.
It shows that the errors $\tilde t_{k}^{(i)}$ for all $i\in N$ can be fast reduced to a small fraction of $d$ with not much large $k$ in all the simulated gird networks.
Meanwhile, in Fig.~\ref{fig:sim_natural_gird3_pattern}, the \emph{source} pattern of the last o-MEP in the sp-MEP process simulated in Fig.~\ref{fig:sim_natural_gird3} is also shown.
We can see that almost all cells of the $256$-cell network become \emph{source} cells with sufficiently large $k$.

\begin{figure}[htbp]
\centering
\begin{subfigure}{.23\textwidth}
\centering\includegraphics[width=1.8in]{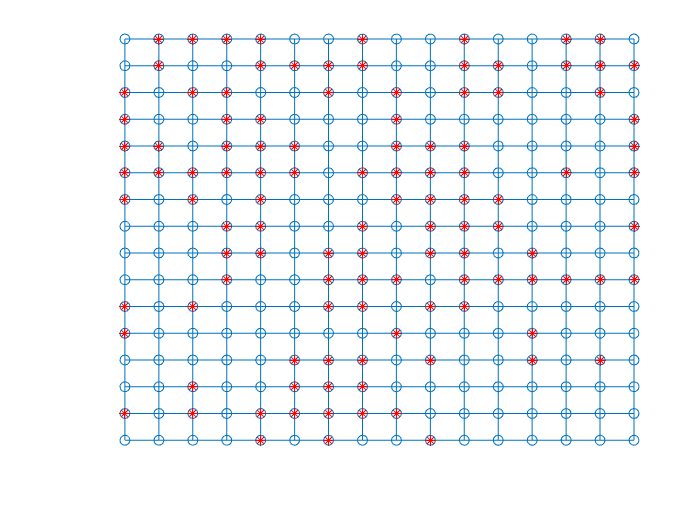}
\caption{with $k=5$}
\label{fig:sim_propa_gird0}
\end{subfigure}
\begin{subfigure}{.23\textwidth}
\centering\includegraphics[width=1.8in]{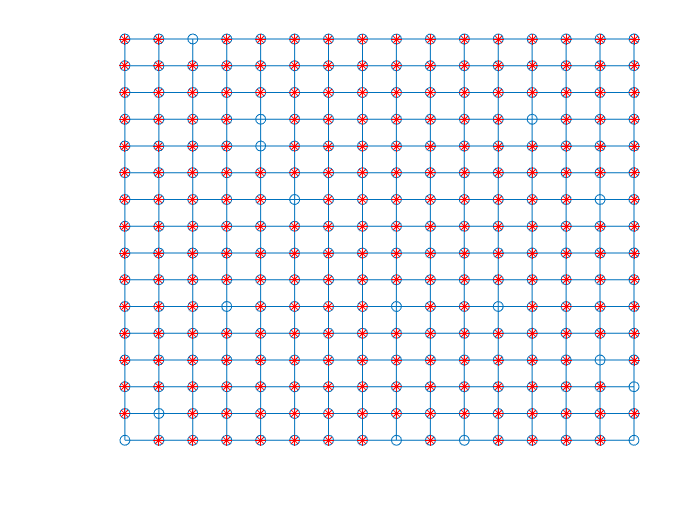}
\caption{with $k=100$}
\label{fig:sim_propa_gird1}
\end{subfigure}
\begin{subfigure}{.23\textwidth}
\centering\includegraphics[width=1.8in]{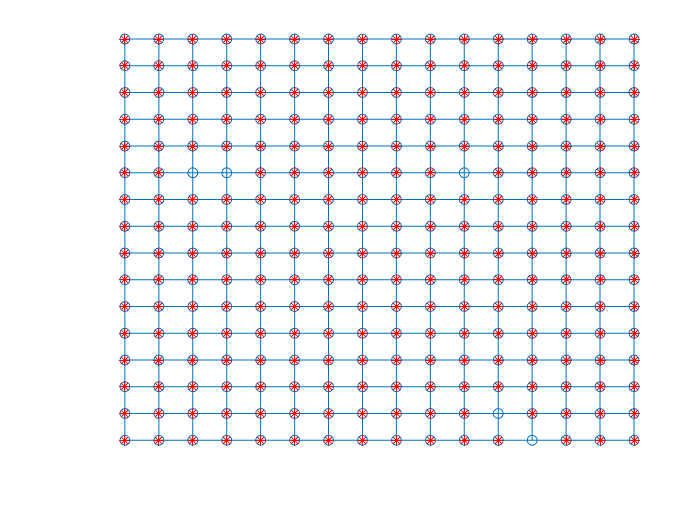}
\caption{with $k=400$}
\label{fig:sim_propa_gird2}
\end{subfigure}
\begin{subfigure}{.23\textwidth}
\centering\includegraphics[width=1.8in]{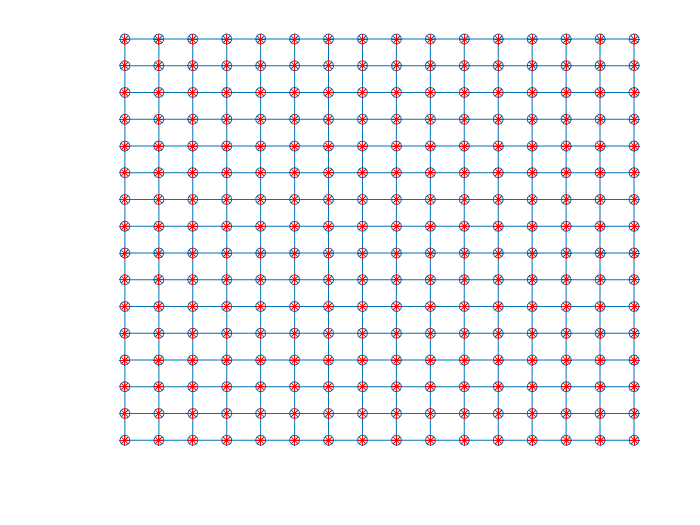}
\caption{with $k=1600$}
\label{fig:sim_propa_gird3}
\end{subfigure}
\caption{Changes of the \emph{source} patterns}
\label{fig:sim_propa_gird}
\end{figure}

Next, in Fig.~\ref{fig:sim_propa_gird}, the changes of \emph{source} patterns of the o-MEPs during the simulated sp-MEP process in the $256$-cell gird are also shown.
We can see that, with the increase of $k$, the number of the \emph{source} cells in the $k$th o-MEP becomes larger and larger until $N_{\text{source}}=N$ holds with a high probability.
Also, as long as $N_{\text{source}}(\mathcal{P}(k))\ne N$, we have $e_1(k+1)< e_1(k)$.
Obviously, for any o-MEP $\mathcal{P}(k)$, when $N_{\text{source}}=N$, we have $N_{\text{bank}}=N_{\text{united}}=N$ and $N_{\text{flow}}=N_{\text{flat}}=N_{\text{ridge}} =\emptyset$.
For convenience, we say the propagation pattern is ideal iff $N_{\text{source}}=N$.

\begin{figure}[htbp]
\centering
\begin{subfigure}{.23\textwidth}
\centering\includegraphics[width=1.8in]{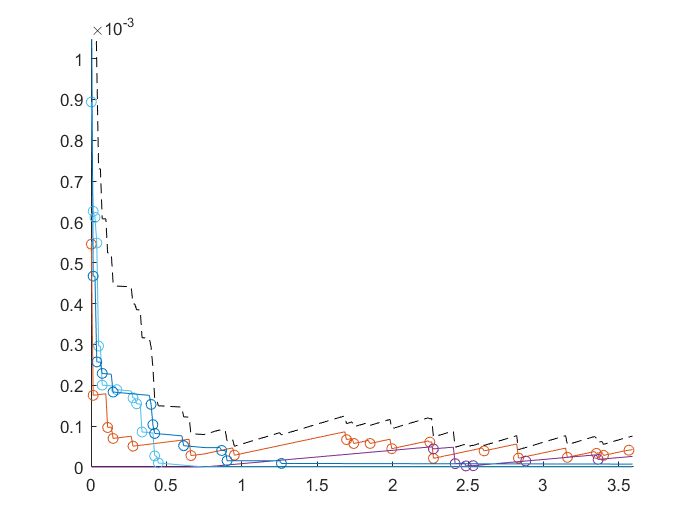}
\caption{for a $4$-cell gird}
\label{fig:sim_natural_rho_gird1}
\end{subfigure}
\begin{subfigure}{.23\textwidth}
\centering\includegraphics[width=1.8in]{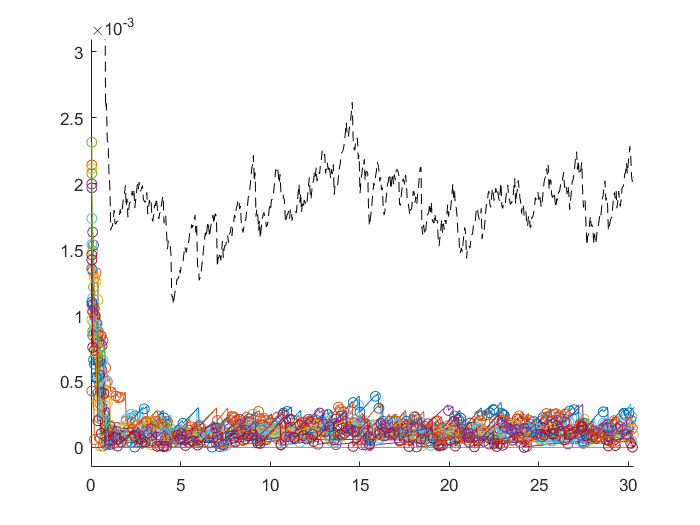}
\caption{for a $16$-cell gird}
\label{fig:sim_natural_rho_gird2}
\end{subfigure}
\begin{subfigure}{.23\textwidth}
\centering\includegraphics[width=1.8in]{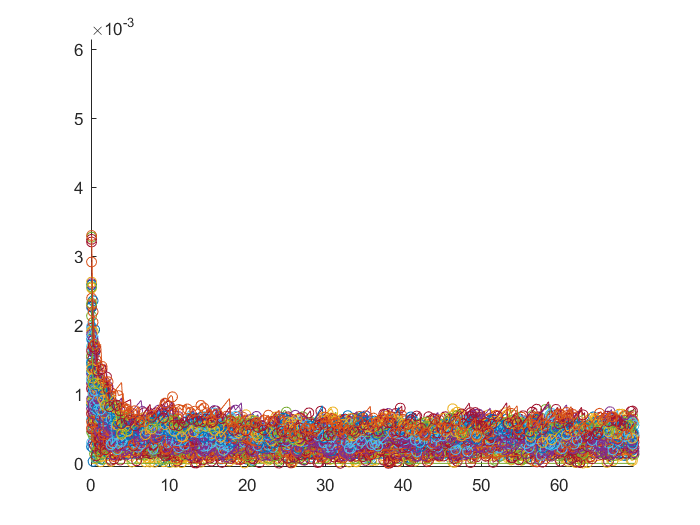}
\caption{for a $64$-cell gird}
\label{fig:sim_natural_rho_gird3}
\end{subfigure}
\begin{subfigure}{.23\textwidth}
\centering\includegraphics[width=1.8in]{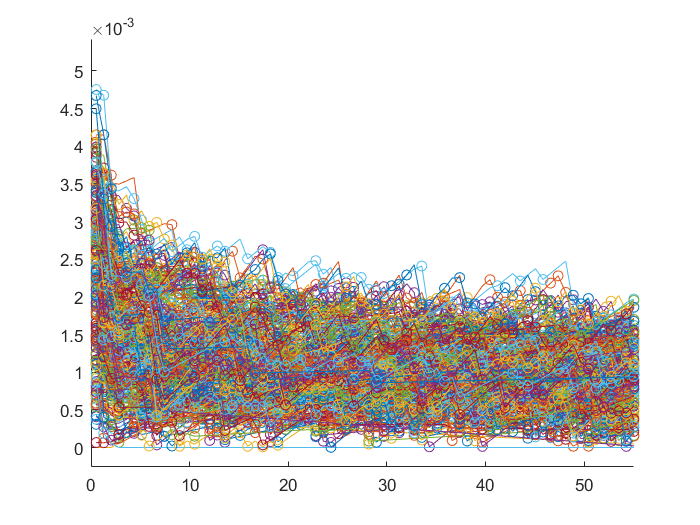}
\caption{for a $256$-cell gird}
\label{fig:sim_natural_rho_gird4}
\end{subfigure}
\caption{The sp-MEP processes with $\rho=10^{-4}$}
\label{fig:sim_natural_rho_gird}
\end{figure}

Then, in Fig.~\ref{fig:sim_natural_rho_gird}, we run the simulation with the same system settings as in Fig.~\ref{fig:sim_natural_gird} except for $\rho=10^{-4}$.
In this case, as the clock drifts accumulated between the adjacent o-MEPs are approximately $\rho\tau_2$, the propagation errors can hardly be better than $\rho\tau_2$.
Nevertheless, the average propagation errors can still be far less than $L_G d$ and $D_G d$.

\begin{figure}[htbp]
\centering
\begin{subfigure}{.23\textwidth}
\centering\includegraphics[width=1.8in]{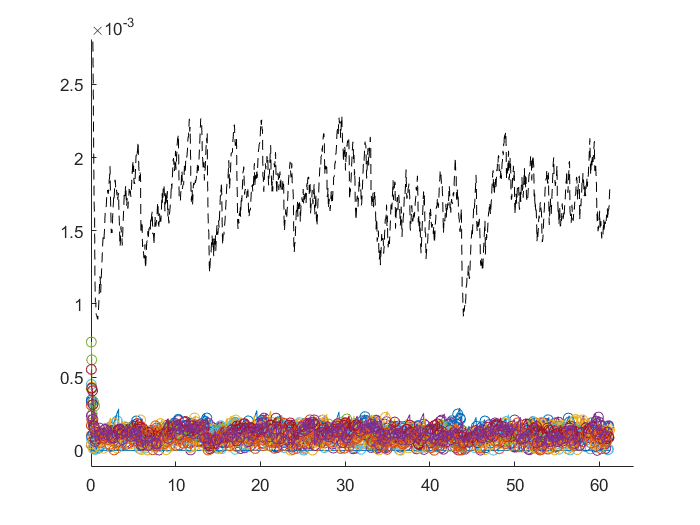}
\caption{for a $16$-cell hypercube}
\label{fig:sim_natural_rho_cube1}
\end{subfigure}
\begin{subfigure}{.23\textwidth}
\centering\includegraphics[width=1.8in]{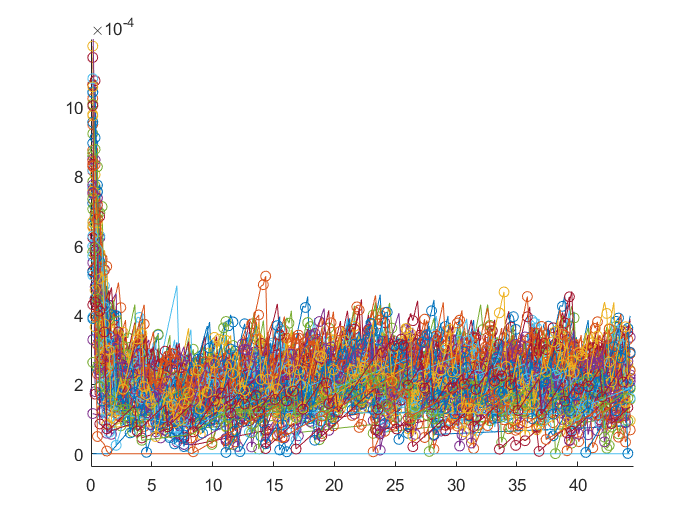}
\caption{for a $64$-cell hypercube}
\label{fig:sim_natural_rho_cube2}
\end{subfigure}
\begin{subfigure}{.23\textwidth}
\centering\includegraphics[width=1.8in]{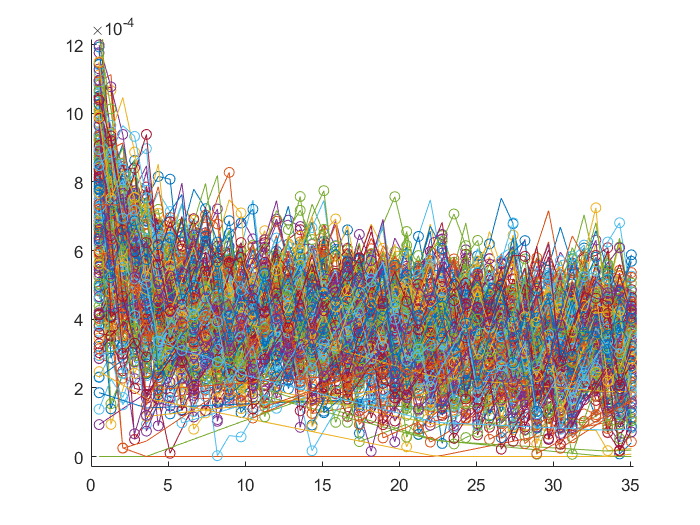}
\caption{for a $256$-cell hypercube}
\label{fig:sim_natural_rho_cube3}
\end{subfigure}
\begin{subfigure}{.23\textwidth}
\centering\includegraphics[width=1.8in]{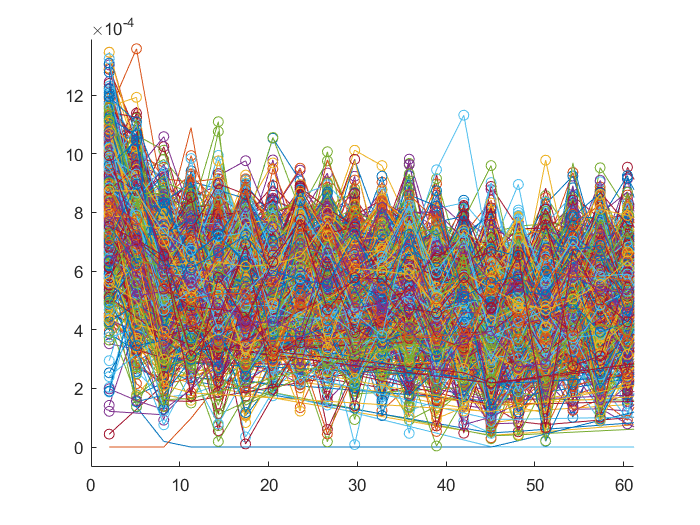}
\caption{for a $1024$-cell hypercube}
\label{fig:sim_natural_rho_cube4}
\end{subfigure}
\caption{The sp-MEP processes in hypercubes}
\label{fig:sim_natural_rho_hypercube}
\end{figure}

Lastly, in Fig.~\ref{fig:sim_natural_rho_hypercube}, we run the simulation with the same system settings as in Fig.~\ref{fig:sim_natural_rho_gird} except for the network topologies being hypercubes (regular graphs with logarithmic node-degrees).
It shows that the MEP systems running upon hypercubes can fast converge compared to the ones upon the girds and rings.
This is mainly because the diameter of a hypercube is logarithmic to the number of cells.

The above discussion only handles the cases with the minimal signal delay $d_{min}=0$.
Practically, for $d_{min}>0$, as the natural timing adjustments of the cells rely only on the internal trigger events, the convergence of the MEP system would become slower.
Also, the propagation errors cannot be further reduced when the trigger instants of the adjacent cells are $d_{min}$ apart in an o-MEP.
In this case, we can add some simple operations in the cells for optimizing the propagation results.
Firstly, as an internally triggered cell $i\in N$ knows that it would not be a source cell in the current o-MEP, $i$ can further adjust its timer (to be faster) with $d_{min}$.
Namely, every cell $i$ can reduce its elapsed local time by $d_{min}$ after $i$ being internally triggered.
With this, the impact of the nonzero $d_{min}$ can be mitigated to some extent.
However, the propagation errors cannot be further reduced with ideal propagation pattern merely with this simple additional operation.
When $d_{min}$ is large, ideal propagation patterns do not necessarily mean small propagation errors.
In this case, instead of directly employing the global broadcast strategy taken in \cite{YuCOTS2021}, localized algorithms with localized report messages can be further developed in reducing the propagation errors.
But here, we limit ourselves with only the one-bit messages.

\subsection{For fault tolerance}
\begin{figure}[htbp]
\centering
\begin{subfigure}{.23\textwidth}
\centering\includegraphics[width=1.8in]{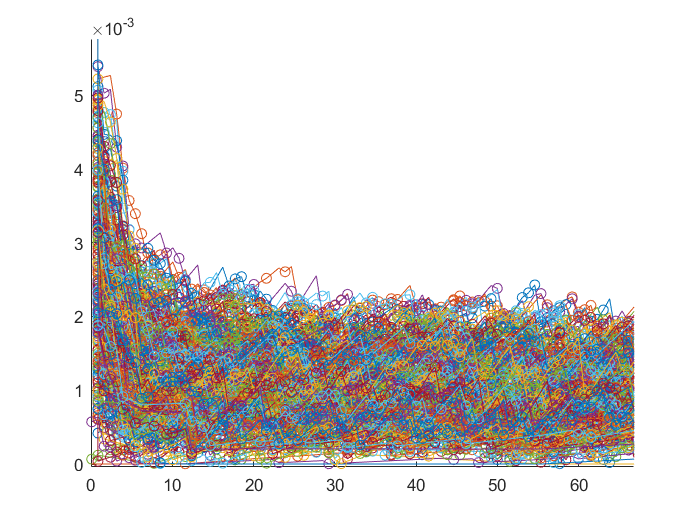}
\caption{with $p=0.01$}
\label{fig:sim_natural_rho_gird_p0}
\end{subfigure}
\begin{subfigure}{.23\textwidth}
\centering\includegraphics[width=1.8in]{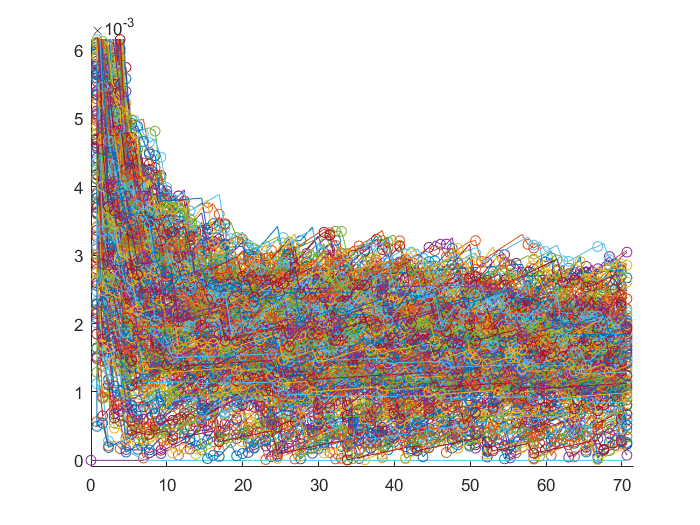}
\caption{with $p=0.1$}
\label{fig:sim_natural_rho_gird_p1}
\end{subfigure}
\begin{subfigure}{.23\textwidth}
\centering\includegraphics[width=1.8in]{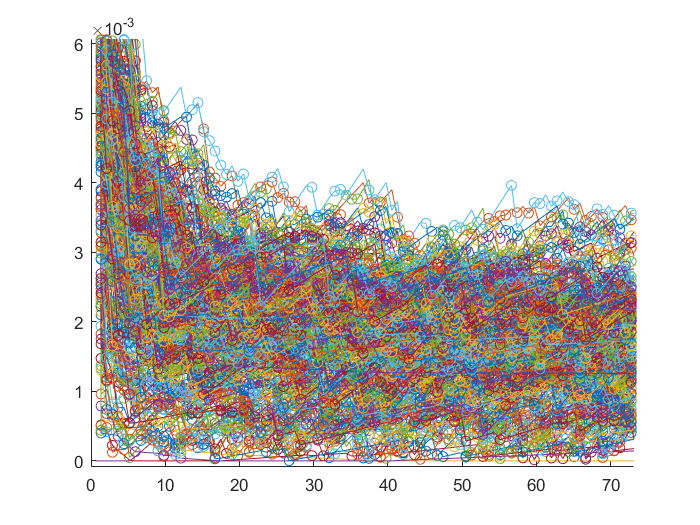}
\caption{with $p=0.2$}
\label{fig:sim_natural_rho_gird_p2}
\end{subfigure}
\begin{subfigure}{.23\textwidth}
\centering\includegraphics[width=1.8in]{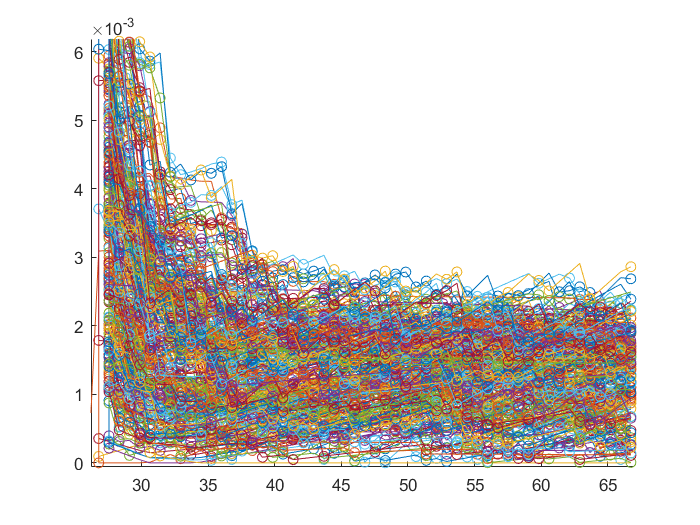}
\caption{with $p=0.3$}
\label{fig:sim_natural_rho_gird_p3}
\end{subfigure}
\caption{The sp-MEP processes with $\rho=10^{-4}$}
\label{fig:sim_natural_rho_gird_faulty}
\end{figure}

Besides optimizing the propagation errors, another practical consideration of the MEP systems is fault tolerance.
Firstly, some benign faults (such as the fail-silent and the fail-omission ones) occurred in cells $F\subset N$ can be naturally tolerated in the MEP systems as long as the remaining correct cells $Q=N\setminus F$ are still connected in $G'=(Q, E\cap(Q\times Q))$ and $L_{G'}$ is still within the range with corresponding to $\tau_0$.
In Fig.~\ref{fig:sim_natural_rho_gird_faulty}, we run the simulation with the same system settings as in Fig.~\ref{fig:sim_natural_rho_gird4} except for allowing some small probability $p$ that each cell $i\in N$ would not be internally triggered when it should be in satisfying the propagation rule described in (\ref{eq:sig_propa}).
It shows that, with the increase of $p$ (being not larger than $0.3$), although the stabilization of the system becomes slower, $\mathcal{S}_0$ can still be stabilized in a sufficiently long time.
However, when we continue to enlarge $p$ beyond $0.3$ in the $256$-cell gird network, the system can remain in unstabilized states for a long simulation time.

In considering malign faults, the situation is not much in favor of the basic MEP system described with (\ref{eq:sig_propa}).
For example, when we consider only the faulty cells with over-frequently emitted trigger signals~\citep{Daniel2018Babbling}, the MEP processes can be significantly affected.
In this case, the trigger events of the correct cells would be affected by the randomly emitted trigger signals of the faulty cell, and thus the desired p-MEP processes cannot be reached.
In handling this problem, there can be several approaches.
Firstly, as the trigger signals emitted from faulty and correct cells can hardly be differentiated with just $1$-bit messages in the sparsely connected networks, we can extend the MEP systems with multi-bits messages or some better-connected networks to isolate the faulty cells.
However, the cost of realizing such kind of self-stabilizing MEP systems would be significantly increased.
Secondly, we can utilize the basic MEP processes as low-layer primitives for constructing high-layer fault-tolerant protocols with multiple MEP subnetworks.
Last but not least, some lightweight operations can be further explored in handling the faults that lay between the benign and the Byzantine ones.
For example, the MEP systems can be extended to satisfy the R\ref{enum_r3} rule with some stricter excitation threshold than (\ref{eq:sig_propa}).
But this is beyond the scope of the paper.

\section{Conclusion}\label{sec:Conclusion}
In this paper, we have investigated the self-stabilizing periodic MEP problem with a self-contained discrete system model.
Inspired by the reaction-diffusion processes, the self-stabilizing periodic MEP processes are intuitively discussed and formally analyzed.
It shows that, just like the desired mechanical power can be periodically outputted by the combustion engines, the desired well-separated trigger events can be periodically outputted by the MEP systems in a self-stabilizing way.
For this, we have shown that by configuring the local restoration threshold $\tau_0$ being approximately $(L_G+1)d$ and the local liveness threshold $\tau_2$ being approximately $3\tau_0$, the $(\tau_\pi,\tau_\Delta,\tau_\nabla)$ sp-MEP system can be deterministically stabilized in $O(L_G)$ time with arbitrary initial states.
Meanwhile, with an extended discussion of the propagation patterns with randomly distributed signal delays, we have shown how the proposed MEP systems can reduce the propagation errors in a natural way.
For this, numeric simulations have been performed upon some specific sparsely connected networks to show the MEP systems' average properties.
Some simple local operations and fail-omission faults are also covered in showing the applicability of the basic MEP primitives in real-world systems.
Featured in the very simple system model, the realization of the proposed MEP primitives can be pretty easy.
With this, the self-stabilizing periodic MEP processes can be employed as low-layer propagation primitives in building easy-understood high-layer self-stabilizing protocols in sparsely connected networks of large scales.

Despite the merits, one significant disadvantage of the basic MEP systems is their vulnerability to malign faults.
Although randomly distributed fail-silent and fail-omission faults can be well tolerated in the basic MEP systems, the MEP processes can be significantly affected by over-frequently emitted trigger signals.
For reaching better fault-tolerance with the proposed MEP systems, several possible approaches are mentioned while the good balance between the simpleness and the robustness of the MEP systems is still an ongoing work.

\bibliographystyle{IEEEtran}
\bibliography{IEEEabrv,MEP}

\end{document}